\documentclass[twocolumn,prb,amsmath,amssymb,amsfonts,superscriptaddress,floatfix,showpacs]{revtex4}

\usepackage{graphicx}
\usepackage{dcolumn}
\usepackage{bm}
\usepackage{rotating} 

\begin{document}

\title{ Pressure Induced Valence Transitions in $f$-Electron Systems}

\author{W. M. Temmerman}
\affiliation{Daresbury Laboratory, Daresbury, Warrington WA4 4AD, UK}
\author{A. Svane}
\affiliation{Department of Physics and Astronomy, University of Aarhus, DK-8000 Aarhus C, Denmark}
\author{L. Petit}
\affiliation{Computer Science and Mathematics Division, and Center for Computational 
 Sciences, Oak Ridge National Laboratory, Oak Ridge, TN 37831, USA}
\author{M. L\"uders}
\affiliation{Daresbury Laboratory, Daresbury, Warrington WA4 4AD, UK}
\author{P. Strange}
\affiliation{School of Physical Sciences, University of Kent, Canterbury, Kent, CT2 7NH, UK}
\author{Z. Szotek}
\affiliation{Daresbury Laboratory, Daresbury, Warrington WA4 4AD, UK}

\date{\today}

\begin{abstract}
A review is given of pressure induced valence transitions in $f$-electron systems
calculated with the self-interaction corrected local spin density (SIC-LSD) approximation.
These calculations show that the SIC-LSD is able to describe
valence changes as a function of pressure or chemical composition.
An important finding is the dual character of the $f$-electrons
as either localized or band-like.
A finite temperature generalisation is presented
and applied to the study of the p-T phase diagram of the $\alpha\rightarrow\gamma$ phase transition in Ce.
\end{abstract}

\maketitle

\section{Introduction}

%

The knowledge of valence of rare earth or actinide ions is important for the understanding of
the solid state physical properties of $f$-electron systems. 
It determines for example the equilibrium volume of $f$-electron compounds.
This is extremely well illustrated in work of Jayaraman\cite{Jaya1,Jaya2}. 
In Fig. \ref{sulphides} we show 
the lattice constants for the rare earth sulphides, selenides and tellurides (after Ref. \onlinecite{Jaya2}).
The abrupt expansion of the lattice for SmS, SmSe, SmTe, EuS, EuSe, EuTe, TmSe, TmTe, YbS, YbSe and YBTe is
associated with a change in valence from trivalent rare earth to divalent rare earth.
Hence this figure shows a direct correlation between lattice constants and valence. Whilst the determination of the valence
is important for the static properties of $f$-electron systems, it becomes less useful
for understanding the dynamical properties such as heavy fermion and Kondo screening of local spin magnetic moments.
However it was shown\cite{Lueders} that the construction of an alloy analogy for the valences can describe 
the finite temperature behaviour, in particular the
pressure versus temperature phase diagram of Ce. Therefore dynamical fluctuations between valences
could possibly describe the finite temperature behaviour of such systems as heavy fermions.

The $f$-electron systems are a rich hunting ground for structural and valence transitions 
as a function of pressure.\cite{Shi1,Shi2,Shi3}
Advances in experimental techniques, such as progress in high pressure cell technology, 
make pressure experiments feasible over an ever-increasing range, possibly up to 1000 kbar.
Furthermore the constant improvements in brightness of the synchrotron sources allow
for the study of high pressure phases of $f$-electron materials  in greater detail.

In this paper we will review a methodology for ab initio calculations
of valence and hence valence transitions as a function of pressure.
The methodology is based on the self-interaction corrected local spin density (SIC-LSD) approximation.
SIC-LSD corrects for a spurious self-interaction of an electron with itself.
This self-interaction is in most circumstances small and hence the LSD is sufficiently accurate for most
applications. However for localized states this correction is substantial and the SIC-LSD approach needs 
to be applied. 
The SIC-LSD method differentiates between localized and itinerant electrons 
and leads to an orbital dependent potential. 
Hence, self-interaction corrections capture a dual picture of 
coexisting localized and band-like $f$-electrons.\cite{nature,SSC} 
As a function of volume or pressure rearrangements in the number of localized and band-like $f$-electrons
take place and a study of this forms one of the topics of this paper.


The paper is organized as follows. In Section II the basic aspects of the SIC-LSD method are
outlined, concentrating on its implementation in terms of bands and k-points. In Section
III we comment on the so-called local SIC implementation and its generalization to alloys
and finite temperatures, allowing to study both static valence and spin fluctuations.
Section IV contains the calculations of valence of elemental rare earths, their mono-sulphides and mono-nitrides.  
Also the results for the valences for Ce-, Sm-, Eu- and Yb-mono-pnictides and mono-sulphides are presented
in that Section. Section V discusses the valences of the actinides and their compounds.
The results for the $\alpha\rightarrow\gamma$ phase transition in Ce at finite temperatures is the subject of Section VI.
Section VII presents the summary and conclusion.

\section{The SIC-LSD total energy method and valence}

The total energy functional of the
local spin density (LSD)
approximation to density functional theory provides good  accuracy
in describing conventional solids with weakly correlated electrons.\cite{gunnarsson,martin} 
However, for all rare earths and for materials containing actinide elements (beyond Np)
 the electron correlations are significant,
and 
the self-interaction correction needs to be included
to obtain a similar accurate description of the localized nature of the $4f$ or $5f$ electrons.

The ensuing SIC-LSD total energy functional\cite{zp} is derived from the LSD
as:
\begin{eqnarray}
\label{Esic}
E^{SIC-LSD}&=&E^{LSD}-\Delta E_{sic},\\
E^{LSD}&=&T+U+E_{xc}+V_{ext}+E_{so},\\
\Delta E_{sic}&=&\sum_{\alpha }^{occ.}\delta _{\alpha }^{SIC}, \\
E_{so}&=&\sum_{\alpha }^{occ.}\epsilon_{\alpha}^{so}
\end{eqnarray}

\noindent where $\alpha $ labels the occupied electron states and $\delta _{\alpha }^{SIC}$ is
the self-interaction correction for state $\alpha $. As usual, $E^{LSD}$ can
be decomposed into a kinetic energy, $T$, a Hartree energy, $U$, the
interaction energy with the atomic ions, $V_{ext}$, and the exchange and
correlation energy, $E_{xc}$.\cite{kohn} 
Here in addition we include the spin-orbit coupling term $E_{so}$.
The spin-orbit energy, for each occupied state $\alpha$ is:
\begin{equation}
\label{Eso}
\epsilon_{\alpha}^{so}= \langle \psi_{\alpha} | \xi(\vec{r})\vec{l}\cdot\vec{s} | \psi_{\alpha} \rangle .
\end{equation}
The $\Delta E_{sic}$ term expresses the spurious self-interaction of the LSD energy, which is 
subtracted from the LSD energy in equation (1). The self-interaction is calculated
for each occupied state $\alpha $, and is 
given by
the sum of the Hartree interaction and the exchange-correlation energy for
the charge density of that state: 
\begin{equation}
\label{dsic}
\delta _{\alpha }^{SIC}=U[n_{\alpha }]+E_{xc}[n_{\alpha }].
\end{equation}
For itinerant states, $\delta _{\alpha }^{SIC}$ vanishes identically, while
for localized (atomic-like) states the self-interaction may be appreciable.
Thus, the self-interaction correction constitutes a negative energy
contribution gained by an $f$-electron when localizing, 
which competes with the band formation energy gained by the
$f$-electron if allowed to delocalize and hybridize with the available conduction states. 
The volume dependence of $\delta_{\alpha}$ is much weaker than the volume dependence of the
band formation energy of rare earth $4f$ or actinide $5f$ electrons, hence
  the overbinding 
of the LSD approximation for narrow $f$ band states is reduced when localization is allowed.
The SIC-LSD energy functional in Eq. (1) appears to be a functional of all the one-electron orbitals,
but can in fact be viewed as a functional of the total (spin) density alone, as discussed in Ref.
\onlinecite{comment}.

One major advantage of the SIC-LSD energy functional 
is that it allows for different valence scenarios to  be
explored. By assuming atomic configurations with different total numbers of
localized states, self-consistent minimization of the total energy leads to different local
minima of the same functional, $E^{SIC-LSD}$ in Eq. (1), 
and hence their total energies may be compared. The
configuration with the lowest energy defines the ground state configuration. Note,
that if no localized states are assumed, $E^{SIC-LSD}$ coincides with the
conventional LSD functional, i.e., the Kohn-Sham minimum of the $E^{LSD}$
functional is also a local minimum of $E^{SIC-LSD}$. 
Among the  actinide elements, this was also
found to be the global minimum for U and Np, but for the later actinide elements, and also in all rare earths,
minima exist with a finite number of localized states.\cite{SSC}
The reason is that the respective $f$ 
orbitals are sufficiently confined in space to benefit appreciably from the
self-interaction correction.

Another advantage of the SIC-LSD scheme is the possibility to localize $f$-states of 
different character. 
In particular the various crystal field eigenstates, either magnetic or paramagnetic.

The SIC-LSD still considers the electronic structure of the solid to be
built from individual one-electron states, but offers an alternative description
to the Bloch picture, namely in terms of
periodic arrays of localized atom-centered states ({\it i.e.}, the
Heitler-London picture in terms of Wannier orbitals). Nevertheless, there
still exist states which will never benefit from the SIC. These states
retain their itinerant character of the Bloch form, and move in the
effective LSD potential. This is the case for the non-$f$ conduction electron states in the
rare earth and actinide metals.

Results of two implementations of the SIC-LSD scheme will be presented in this paper.
The first approach applies the SIC to the localized bands, whilst in the second approach
the SIC is applied to the phase shifts of the localized states. 
Both approaches can be considered to be a reasonable ansatz, for the solid state, to the implementation of the
SIC for atoms.
The latter approach is local in nature - as expressed by the use of single site phase shifts- whilst the former
can take into account a more extended behaviour of the wave function. 
However the band-based approach involves repeated back and forth transformations of
the wavefunctions from reciprocal space to real space localised functions.
In real space the band-dependent SIC potential is evaluated,
in reciprocal space the bandstructure problem is solved.
This is a time consuming aspect of this method.
Further details of this implementation
can be found in Ref. \onlinecite{brisbane}.
For a variety of applications to $d$ and $f$ electron solids, see Refs.
\onlinecite{brisbane,nature,pss,rules,puo2} and references therein.
The results in Sections IV and V
are based on applying the SIC to the localized bands, whilst the finite temperature
discussion of Section VI is based on the single site phase shifts and the so-called L(ocal)-SIC
implementation\cite{Lueders} in terms of multiple scattering theory, briefly outlined in the following section.

\section{Local Self-Interaction Approach}

In this local formulation of SIC we use the multiple scattering formalism
and concentrate our attention on the
phase shifts of electrons scattering from ions in a solid. If a phase
shift is resonant it is reminiscent of a bound state at positive
energies, i.e., above the zero of the potential which in this case is
the muffin-tin zero.  The energy derivative of the phase shift is
related to the Wigner delay-time.  If this is large the electron will
spend a long time on the site.  Such 'slow' electrons will be much
more affected by the spurious self-interaction and therefore should
see an SI-corrected potential.
Thus when the phase shift has a resonance we calculate the
self-interaction correction in this $(l,m)$ angular momentum channel.
This is accomplished by calculating the one-electron charge
density for this channel, which defines the charge density for the
self-interaction correction potential to be added to the LSD potential.
Then the phase shifts using the total (SIC-LSD) potential are recalculated.
This is applied $m$ channel by $m$ channel for a particular angular
momentum $l$. Like in the case of band by band SIC implementation,
the minimization of the total energy determines the
optimum configuration of ($l$,$m$) channels for self-interaction
correction. Therefore we can associate with each of the $m$ channels
two potential functions, V$^{\rm SIC-LSD}_{\rm eff}$(r) and
V$_{\rm eff}^{\rm LSD}$(r). If the total energies of these scenarios are
sufficiently close, one can envisage dynamical effects playing an
important role as a consequence of tunneling between these states.

Due to the multiple scattering aspect of this approach we can easily
calculate Green's functions and from them various observables for
making contact with experiments.
Another advantage is that it can be easily generalized to include the
coherent potential approximation, \cite{Soven:67, StocksEtAl:78, GyorffyStocks:79,
FaulknerStocks:81, StocksWinter:84}
extending the range of applications to
random alloys. In addition, one can use it to treat static correlations
beyond LSD by studying pseudoalloys whose constituents are composed of e.g.
two different states of a given system: one delocalized, described by
the LSD potential, and another localized, corresponding to the SIC-LSD
potential.
Combined with the disordered local moments (DLM) formalism for
spin-fluctuations,\cite{DLM1,DLM2} this allows also for different
orientations of the local moments of the constituents involved.
In addition, in this formulation we can generalize the SIC approach
to finite temperature, T, to study its effect on the electronic total
energies, E$_{tot}$, and the electronic contribution to the entropy.

To fully take into account the finite temperature effects, we calculate
the free energy of a (pseudo)alloy, as a function of temperature, volume, 
V, and concentration, c, namely
\begin{eqnarray}
F(T,c,V) &=& E_{\rm tot}(T,c,V)  - T \Big( S_{\rm el}(T,c,V) \nonumber \\
&& + S_{\rm mix}(c)  + S_{\rm mag}(c) + S_{\rm vib}(c) \Big) .
\end{eqnarray}
Here $S_{\rm el}$ is the electronic (particle-hole) entropy, $S_{\rm mix}$
the mixing entropy of the pseudoalloy, $S_{\rm mag}$ the magnetic
entropy, and $S_{\rm vib}$ the entropy originating from the lattice
vibrations.

The full extent of the L-SIC approach is ilustrated in Section VI
where we discuss the famous Ce $\alpha$ $\rightarrow$ $\gamma$ phase
transition.

\section{Localization/delocalization and valence transitions in rare earth compounds}

\subsection{Valencies of rare earths, rare earth nitrides and rare earth sulphides}
\begin{figure}
\begin{center}
\includegraphics[width=90mm,clip]{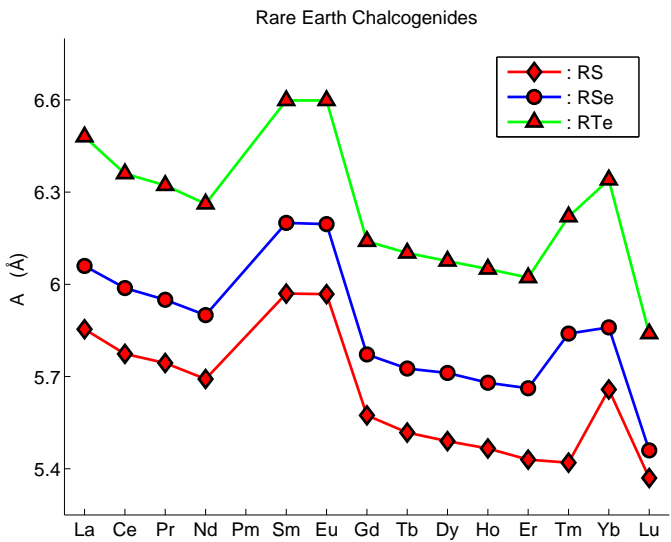}\\
\caption{ 
\label{sulphides}
Lattice constants of rare earth sulphides (after Jayaraman in Reference \onlinecite{Jaya2}).
}
\end{center}
\end{figure}

\begin{figure}
\begin{center}
\includegraphics[width=90mm,clip]{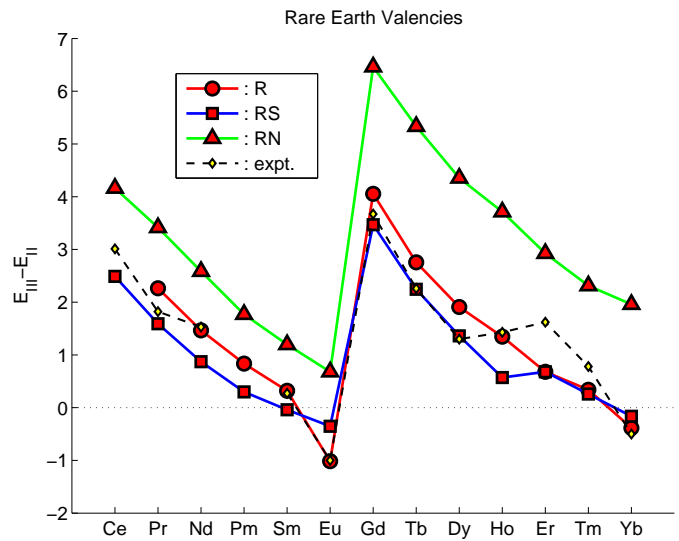}\\
\caption{ 
\label{nature1}
Energy differences (in eV) between the divalent and trivalent configurations 
for the rare earths, their sulphides and nitrides.\cite{nature,not_nature}
The dashed line shows the 'experimental' values for the rare earth metals.\cite{Borje}
The circles, squares and triangles show the calculated values
for the rare earth metals, the rare earth sulphides and the rare earth nitrides, respectively.
The open circles and the crosses show the calculated values
for the rare earth metals and the rare earth sulphides, respectively.
}
\end{center}
\end{figure}

The change of the rare earth valence as a function of atomic number is one of the outstanding physical properties of
elemental rare earths.
Among the materials that exhibit valence changes are the rare earth elements and their sulphides 
where changes from trivalent to divalent are observed in SmS, Eu, EuS, Yb and YbS,
whilst no valence transition is observed in the rare earth nitides.
The lattice constants of all pnictides, with the exception of Ce, behave continuously.  
In the sulphides (Fig. \ref{sulphides}) changes
of the lattice constants as large as 10\% occur in the middle and at the end of the series. 
These are 
associated with a change in valence from trivalent to divalent in the sulphides whilst the pnictides remain trivalent. 
The discontinuous behaviour
of the 
CeN lattice contant is due to tetravalency. 
The calculated SIC-LSD energy differences between divalent and trivalent 
configurations are presented in Fig. \ref{nature1}. 
One sees that at the beginning of the series the rare earths, their nitrides and sulphides are
very solidly trivalent. This tendency for trivalency decreases towards the middle
of the rare earth series where the curve more or less repeats itself from Gd onwards. The nitrides
remain trivalent throughout the rare earth series whereas SmS, Eu, EuS, Yb and YbS become divalent.
From Fig. \ref{nature1} it is clear that the rare earth nitrides are most trivalent and
the rare earth sulphides are least trivalent. The elemental rare earths are intermediate
between the nitrides and sulphides.
Whilst the trends are well reproduced, there is an overall tendency to overemphasize divalency.
Thus all the results n Fig. \ref{nature1} have been calibrated by 43 mRy to agree with the observed valence
transition pressure of 6 kbar in SmS.\cite{Jaya2}

\begin{figure}
\begin{center}
\includegraphics[scale=0.5]{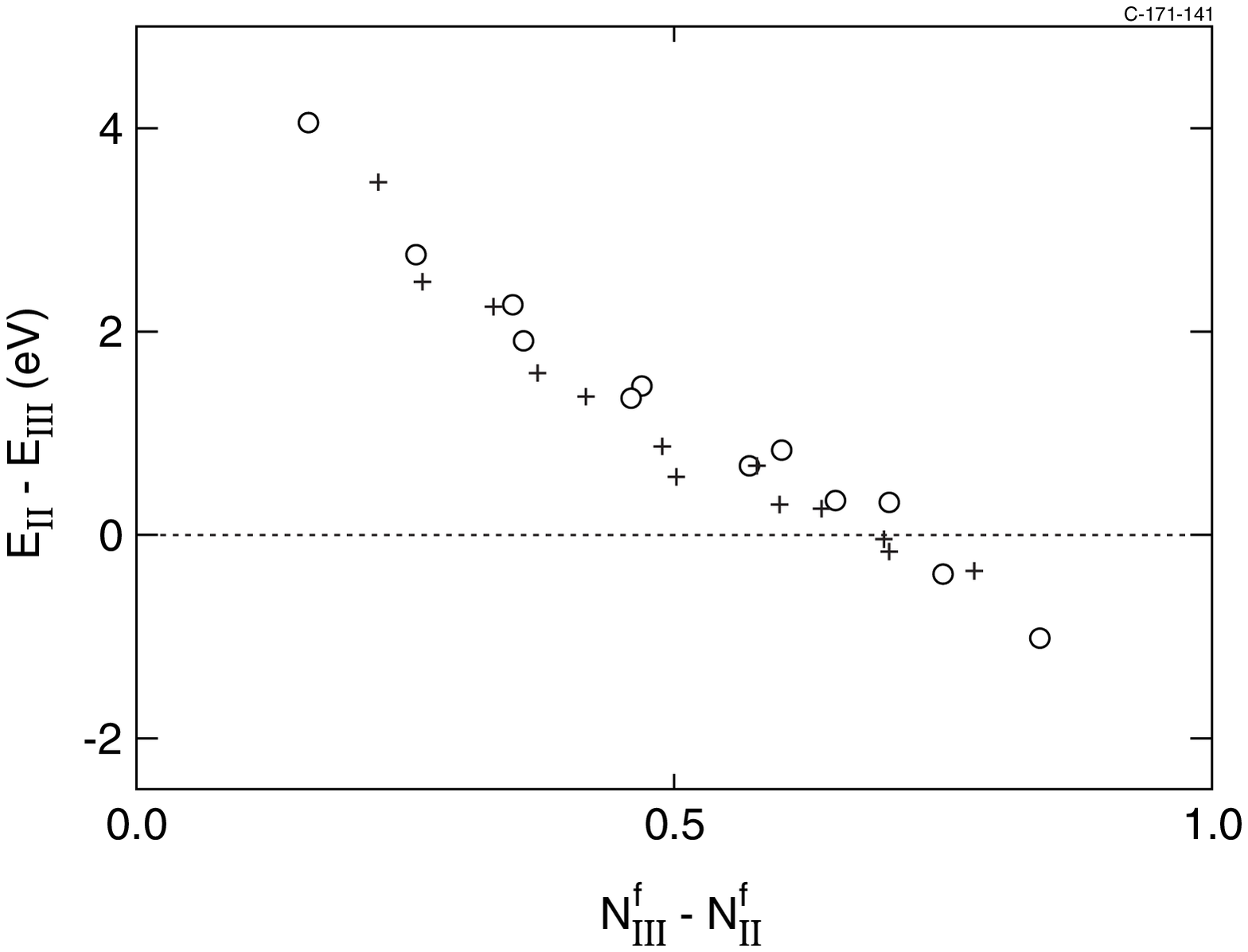}\\
\caption{ 
\label{bandelectrons}
The correlation between rare earth valence and the number of band-like $f$-electrons.
Positive values on the y-axis mean trivalency. Negative values on the y-axis mean divalency.
Note that for more than 0.7 $f$-band electrons the divalent configuration becomes more favourable.
}
\end{center}
\end{figure}

Having firmly demonstrated that the behaviour of the lattice constants seen in 
Fig. \ref{sulphides} is caused by the valence behaviour, 
we can endeavour to ask why.  
In Fig. \ref{bandelectrons} the number of $f$-band electrons in the trivalent 
states is plotted. The crossover from the trivalent to divalent state occurs when the number of
$f$-band electrons is above 0.7. At this point the localization energy wins over the
band formation energy and it becomes more advantageous to localize an extra $f$ electron. 
For the nitrides the number of occupied itinerant $f$-electron states stays well below
0.7 due to the strong hybridization with the nitrogen $p$-states. The nitrides remain
trivalent throughout the rare earth series.
With respect to this the application of the SIC-LSD is different between
atoms and solids. In the case of atoms the levels are occupied in steps of one electron and
the self-interaction is always applied. In the solid state the occupancy of a state can
vary between zero and one and whether or not to apply the self-interaction is determined
by a competition between the band formation energy (no self-interaction) and 
the localization (application of the self interaction) of each of the bands. 
This leads to a definition of valence
since the self interaction corrected localized states are well separated from the valence
states and no longer available for bonding and band formation.     
In the following we will discuss some further applications for the mono-pnictides
and mono-sulphides of Ce, Sm, Eu and Yb.


\subsection{Ce monopnictides and monochalcogenides}

Cerium and cerium compounds attract considerable attention due to the
intricate electronic properties related to the Ce $f$-electrons.
A variety of phenomena like heavy-fermion,\cite{stewart,fulde}
mixed-valence\cite{falicov} and Kondo behaviour\cite{aeppli,kvc}
is encountered.
The cerium  monopnictides and monochalcogenides have peculiar
properties as a function of applied hydrostatic pressure (see Table \ref{Table1}). 

With the SIC-LSD approach\cite{pss}, we have studied the high pressure
behaviour of CeP, CeAs, CeSb, CeBi\cite{ssc,cep}, CeS, CeSe and CeTe\cite{ces}.
For all these compounds we have found the trivalent Ce state to be the
groundstate.
The experimental lattice constants are
reproduced within $1\%$ accuracy.
The results regarding pressure transitions are summarized in Table \ref{Table1}, which 
contains all theoretically computed phase transition pressures and relative volumes
on the low and high pressure sides of the transition (given as a fraction of the
equilibrium volume at $P=0$). 

\begin{figure}
\begin{center}
\includegraphics[width=90mm,clip]{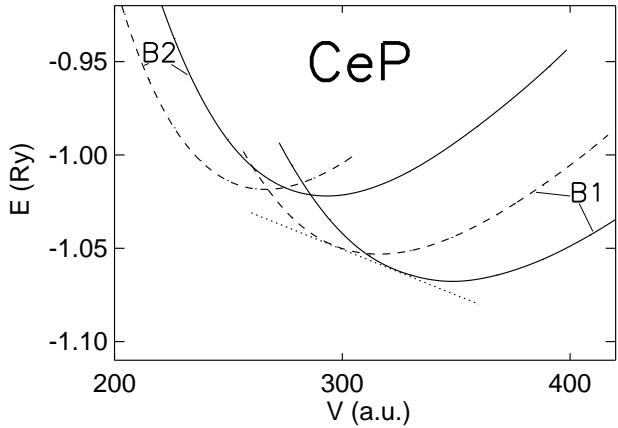}\\
\caption{Cohesive  energy of CeP    (in Ry/formula unit) as a function of
        specific volume (in $a_0^3/$formula unit).
        Two crystal structures, the B1 and the B2,
        are considered, and each with two different treatments of the Ce
        $f$-electrons. The full drawn curves  correspond to
        calculations with one localized $f$-electron per Ce atom, while the
        dashed curves
        correspond to itinerant $f$-electrons.
        The dotted line marks the
        common tangent at the isostructural
        phase transition in the B1 structure.
\label{fig1}}
\end{center}
\end{figure}

   The total energy as a function of volume for CeP is calculated
   for each of the phases B1 and B2 and with the $f$-electron treated as
   either delocalized (normal band picture, as implemented with LSD) 
or localized (SIC-LSD). The results are
   shown in Fig. \ref{fig1}. The lowest energy is found in the B1 phase with
   localized $f$-electrons and
   with a specific volume of $348$ a${}_0^3$
   per formula unit, which coincides with the experimental
   equilibrium volume. 
   The B1 phase with delocalized $f$-electrons has its minimum at a
   considerably lower volume, due to the significant $f$-electron
   band formation energy providing a large negative component
   to the pressure. From the common tangent a phase transition is
   predicted at a pressure of 71 kbar with a volume collapse of
   $\Delta V/V_0= 8\%$
(change in volume relative to the zero pressure equilibrium volume),
   which is in excellent agreement with the transition
seen at 55 kbar.\cite{Mori} The B2 structure is not as favorable
   for the CeP compound, since the calculated energy is substantially
    higher than that of the B1 structure. This holds for both localized
   and delocalized $f$-electrons. From Fig. \ref{fig1} we conclude that the B2
   structure with localized $f$-electrons is never reached in CeP, while at high
   pressure a second phase transition to the B2 structure with
   delocalized $f$-electrons is found.
   The transition pressure is calculated
   to be 113 kbar and the volume collapse 12\%, while experimentally
   the B1$\rightarrow$B2 phase transition is seen at 150$\pm 40$
   kbar.\cite{Vedel} The experimental volume collapse is 11 \%. Some
   uncertainty is associated with comparing the
   total energies calculated for the B1 and B2 phases in the present approach.
   Therefore we have also investigated the high pressure transition in CeP
   with the full-potential LMTO method,\cite{methfessel} where no problem
   of this sort occurs. In this case we have found a transition pressure
   of 167 kbar, which is somewhat higher, but in good agreement with the
   experimental value.

The results reported in Table \ref{Table1} show that all
of the observed pressure transitions in the cerium pnictides and 
chalcogenides are indeed reproduced.\cite{ssc,cep,ces} 
The total energy curves look rather similar to those of 
CeP in Fig. \ref{fig1}, but minor changes in the relative positions occur when the ligand is 
varied. The localized phases are generally more favored when the ligand ion becomes 
heavier, and as a consequence, in CeAs no isostructural delocalization transition occurs in the B1 structure. Instead a transition directly from the B1 structure with localized
$f$-electrons to the B2 structure with delocalized $f$-electrons occurs, in agreement 
with experiment. In CeSb and CeBi the first high pressure transition to occur is from B1 to
B2, with localized $f$-electrons in both cases, and only at higher pressures is a 
delocalization transition predicted to take place. The calculated transition pressures 
are only slightly above the ranges studied experimentally.
In this work only the B2 structure was
considered for the second transition, but in reality the valence 
transition which eventually must take
place in CeSb and CeBi may involve another high pressure phase.

In CeS the first transition occurs to the B1 phase with delocalized
$f$-electrons, {\it i.e.}, the theory predicts an isostructural  phase
transition in CeS. The calculated transition pressure is 101
kbar with a volume collapse of $6\%$.
These findings are in excellent agreement with the experiment of
Ref. \onlinecite{Croft}, but at variance with the results of Ref.
\onlinecite{Vedel-ces}, where no discontinuity in the $pV$-curve is observed.
These results may indicate the proximity of a critical
point.
At higher pressures CeS transforms into the B2 phase. According
to the present calculations this occurs in two steps. First, at a pressure
of 243 kbar, CeS goes into the trivalent B2 phase with a $4.6 \%$  volume
change. In the second step, at a pressure of 295 kbar, the tetravalent B2
phase is reached with a $3.6\%$ volume collapse. Thus, CeS reenters the
localized regime for a very narrow pressure range. Some caution is necessary
before accepting this rather peculiar behaviour. First of all the
present calculations have been performed
 at $T=0$ K, and thermal fluctuations might
easily merge the two high-pressure transitions into one. Secondly, the
SIC-LSD calculational scheme only implements rather idealized pictures of
either completely localized or completely delocalized Ce $f$-electrons. Most
likely, the Ce $f$-electrons enter into a complicated Kondo screened state
in the B2 phase, which would alter the energetics of the B2 phase in a way
we are currently not able to address. One could speculate that the Kondo
screening at $T=0$ K would interpolate between the ideal localized situation at
large volumes and the ideal delocalized situation at small volumes.
Depending on the details of this screening it could turn the 
transition into a continuous valence transition.
Unfortunately, no experiments have
been performed beyond 250 kbar.\cite{Leger}
The almost certain theoretical prediction made here is that at high pressures
the B2 phase will be reached. However, experimental verification is needed.
Since we have found\cite{ssc} that our present calculational approach tends
to underestimate transition pressures for transitions
from the B1 structure to the B2 structure, the
experimental transition pressure for the B1$\rightarrow$B2 transition in CeS
would most likely
be in the range of 300-350 kbar, which is easily within experimental  reach.
Of course, we can not rule out that              other
crystal structures may become important at high     pressures.
In elemental cerium as well
as other rare earths, low-symmetry crystal phases occur when the $f$-electron
delocalization sets in.\cite{mcmahan,nature}

In both CeSe and CeTe the only pressure transition observed is that from B1 to B2
with localized $f$-electrons in both phases. These are also first to occur according
to the calculations, while valence        transitions are predicted in the range of 400 
kbar. Thus, the situation here is quite similar to that in CeSb and CeBi, apart from
the tetragonal distortion in these compounds which was not found for CeSe and 
CeTe.\cite{ces}




\subsection{Sm monopnictides and monochalcogenides}

One of the most studied rare earth compounds is SmS\cite{wachter,Jayaraman,Varma,Varma1}.
At low temperature and zero
pressure it crystallizes in the NaCl structure
exhibiting a semiconducting behaviour.
At a moderate pressure of $\sim 6.5$ kbar  SmS reverts
to a metallic
phase with a significant volume collapse of 13.5\% \cite{Benedict1},
retaining however the
NaCl structure.
Valence transitions can also be
brought about by alloying the SmS lattice with other trivalent ions, such as
Y, La, Ce  or Gd.\cite{wachter,Jayaraman2}
Similar valence instabilities are observed in
SmSe and SmTe \cite{Jayaraman,Campagna,Sidorov,LeBihan},
which  also crystallize in  the NaCl structure.
For these compounds the volume changes continuously, but anomalously,
with pressure  (at room temperature) \cite{Chatterjee,LeBihan}.
From the photoemission studies it is concluded that
SmSe and SmTe at amibient pressure, like SmS,
are also of predominantly  divalent $f^6$  character \cite{Campagna}
while the monopnictides show clear signals of pure trivalent $f^5$ ions \cite{Pollak,Campagna}.

\begin{figure}
\includegraphics[width=.45\textwidth]{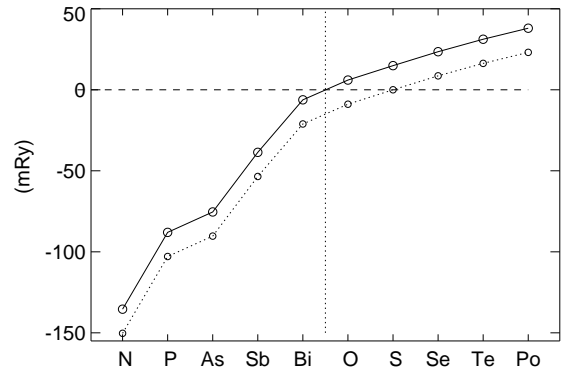}~a)
\hfil
\includegraphics[width=.45\textwidth]{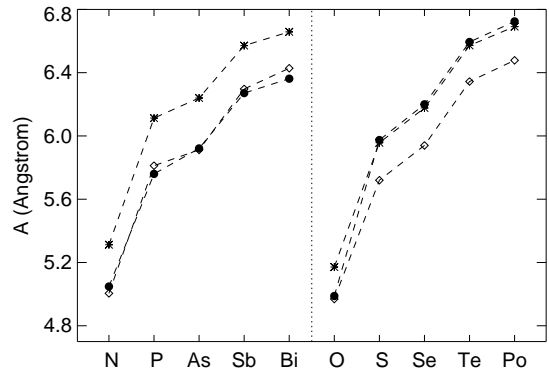}~b)
\caption{a): Trivalent-divalent energy difference, $\Delta E=E(f^5)-E(f^6)$, of samarium compounds.
Positive $\Delta E$ means divalent, negative $\Delta E$ means trivalent.
The dashed line marks the calibrated curve \cite{SmX}. b): Comparison of experimental
and theoretical lattice constants of Sm mono-pnictides and mono-chacogenides compounds.\cite{SmX}
Experimental values are marked with solid circles, while lattice constants calculated
assuming a divalent (trivalent) Sm ion are marked with stars (diamonds).
}
\label{fig2_sm}
\end{figure}

Figure \ref{fig2_sm} shows the calculated lattice constants and valence stabilities for the
Sm pnictides and chalcogenides \cite{SmX,SmX_1}.
Specifically, the total energy difference is calculated for the scenarios of trivalent and
divalent rare earth ions. A positive
energy difference implies that the divalent configuration is preferred.
For the Sm pnicides,
the calculations reveal a strong preference for
the $f^5$  configuration in the early pnictides, with the energy difference of
1.8 eV  per formula unit in SmN. For the heavier Sm pnictides, the $f^6$ configuration
becomes more and more advantageous, and for Bi it is only 0.08 eV higher than the trivalent
configuration. 
This is the same trend towards more localised phases,
as seen in the Ce compounds of the previous section,
when the ligand ion becomes heavier.

Moving to the Sm chalcogenides, already in the Sm monoxide the $f^6$
configuration is found to be most favorable, by 0.08 eV, and in SmS by 0.20 eV.
Hence, the SIC-LSD total energy predicts a valence transition of Sm between the Sm pnictides
and the Sm chalcogenides.
This  is not in complete agreement with   the experimental picture,
according to which the divalent and intermediate-valent states
 are almost degenerate in SmS, while SmO
is trivalent and metallic \cite{smo,Krill}. Thus,  it appears that
the SIC-LSD total energy functional overestimates the tendency to form the
divalent configuration of Sm, by approximately 15 mRy, in SmS.
Assuming a similar error for all Sm compounds,
this would imply that the calculated
energy balance curve in Fig. \ref{fig2_sm} should be lowered by approximately 15 mRy.
The dashed line of the figure shows the energy difference
with such a correction. This
switches the balance in favor of trivalency for SmO,
in accord with experiments \cite{smo,Krill}.
Note that this is a reduction, due to the inclusion of the spin-orbit interaction, 
on the 43 mRy calibration energy quoted earlier and applied in Fig. \ref{nature1}.
In other words, the spin-orbit interaction accounts for 28 mRy of the
43 mRy calibration energy. The remaining 15 mRy is therefore
$f$-$f$ correlations not accounted for by the SIC-LSD.
The lattice constants are seen to be in
excellent agreement with experimental value for all compounds, corroborating the
conclusion that a valence shift occurs between SmO and SmS.

The trivalent phase of the chalcogenides becomes relevant at high pressure. In this
phase the localised $f^5$ Sm ions coexist with a partly occupied
narrow $f$-band, effectively describing an intermediate valent phase \cite{SmX}.
The calculated and measured transition pressures are listed in Table \ref{Ptranssm}.
The good agreement both for transition pressures and volume collapses proves that
the bonding of the high pressure phase is well described in the SIC-LSD
approximation, even if the true many-body wavefunction of the intermediate valence phase
is much more complicated than the corresponding SIC-LSD wavefunction. This is
in line with the general philosophy of the density functional approach of obtaining
good total energy estimates from simple reference systems (non-interacting electrons).
The present theory cannot describe the continuous nature of the transition
observed for SmSe and SmTe. The experiments were all conducted at room temperature and
it would be interesting to investigate whether
the continuous transition would  exist at low temperature
as well.

\subsection{Eu monopnictides and monochalcogenides}

Europium chalcogenides and most of the pnictides crystallise
in the simple NaCl crystal structure and hence form a series that can be
studied within first principles theory relatively easily. Recently the
chalcogenides have attracted a lot of attention due to their potential
applications in spintronic and spin filtering devices.\cite{LeClair}

\begin{figure}
\begin{center}
\includegraphics[scale=0.35,angle=-90]{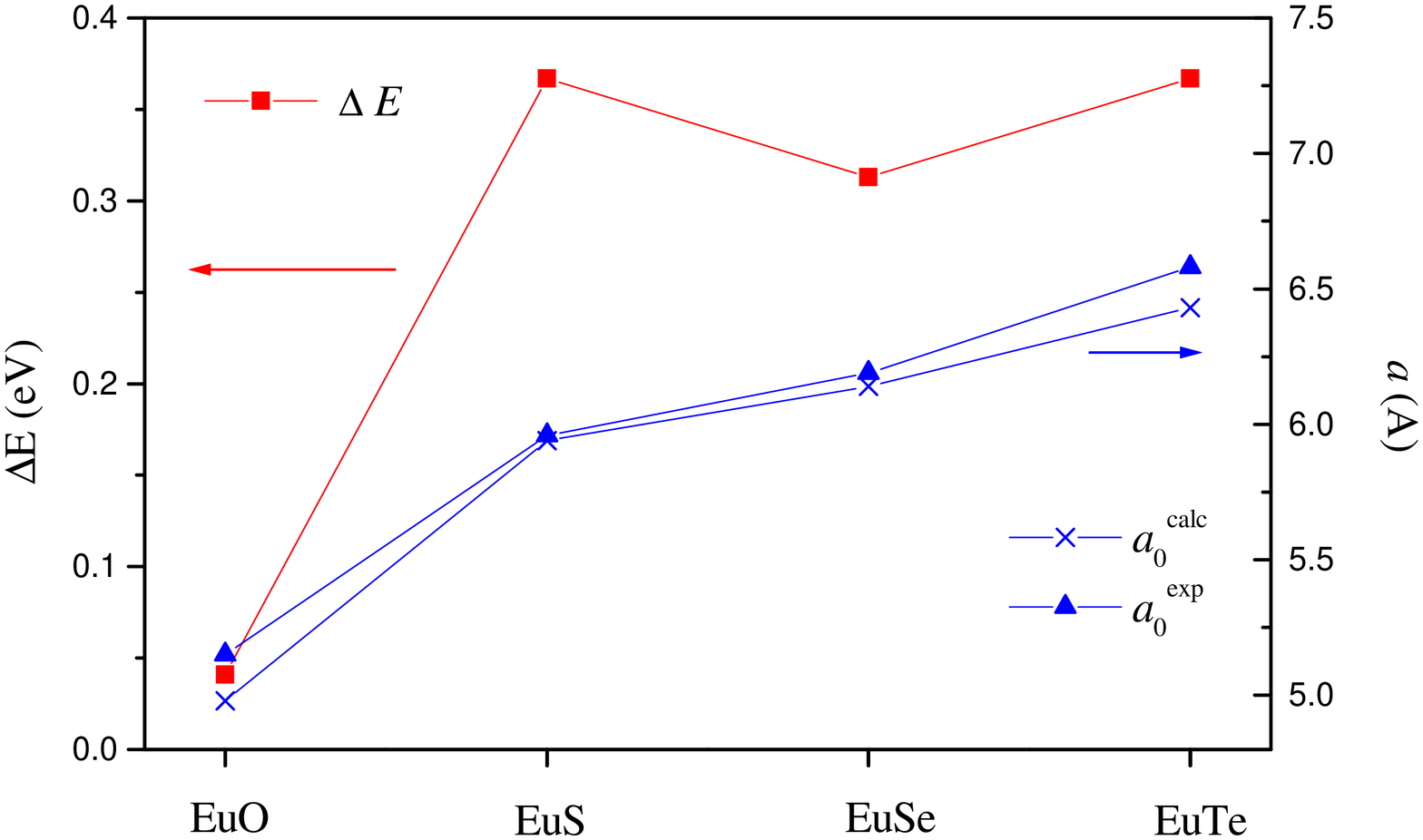}\\
\caption{The values of the lattice constant for the europium chalcogenides
from experiment\cite{gasg} (triangles) and calculation\cite{EuX} (crosses). 
Also shown is the calculated energy difference between the divalent and trivalent
form (squares). 
The positive values mean divalent europium chalcogenides.
\label{fig_EuSulph}
}
\end{center}
\end{figure}

The SIC-LSD method is applied to study the electronic structure
of the Eu compounds in the assumed ferromagnetic state in both the divalent 
and trivalent configurations. Fig. \ref{fig_EuSulph} shows the calculated 
and experimental lattice constants for
all the chalcogenides.\cite{EuX} 
Also shown are the energy differences between the
two valence states. 
It is clear that all the europium chalcogenides are divalent. This is as 
expected from simple shell-filling grounds. It can also be seen from this figure
that the calculated lattice constants are in good agreement with experiment.
The energy difference between the two valence states is fairly independent
of chalcogenide for S, Se and Te and one can observe that there is a clear
correlation between the lattice constant and the difference in energy
between the divalent and trivalent states.

The calculated transition pressures for EuO and EuS are compared
with experiment in Table \ref{Ptranseu}. 
The lower transition pressure and smaller volume in EuS compared to EuO is 
reproduced by the theory.
Experimentally, the transition of EuO
(at room temperature) is continuous, which the present theory cannot describe. 
For EuS the experiments show no anomalous compression curve
\cite{Jayaraman3}, but the band gap closes
at 160 kbar, just before the structural transition to the CsCl structure (at 200 kbar)
\cite{Syassen}. 
However due to the LMTO-ASA approximations,
significant   uncertainty persists in the
values of the total energy differences between different crystal structures.
Also the spin-orbit interaction can significantly alter the results: 
we found that without spin-orbit the structural transition 
occurs at 137 kbar \cite{pss}, however without an isostructural transition occurring
first. 

\begin{figure}
\begin{center}
\includegraphics[scale=0.5]{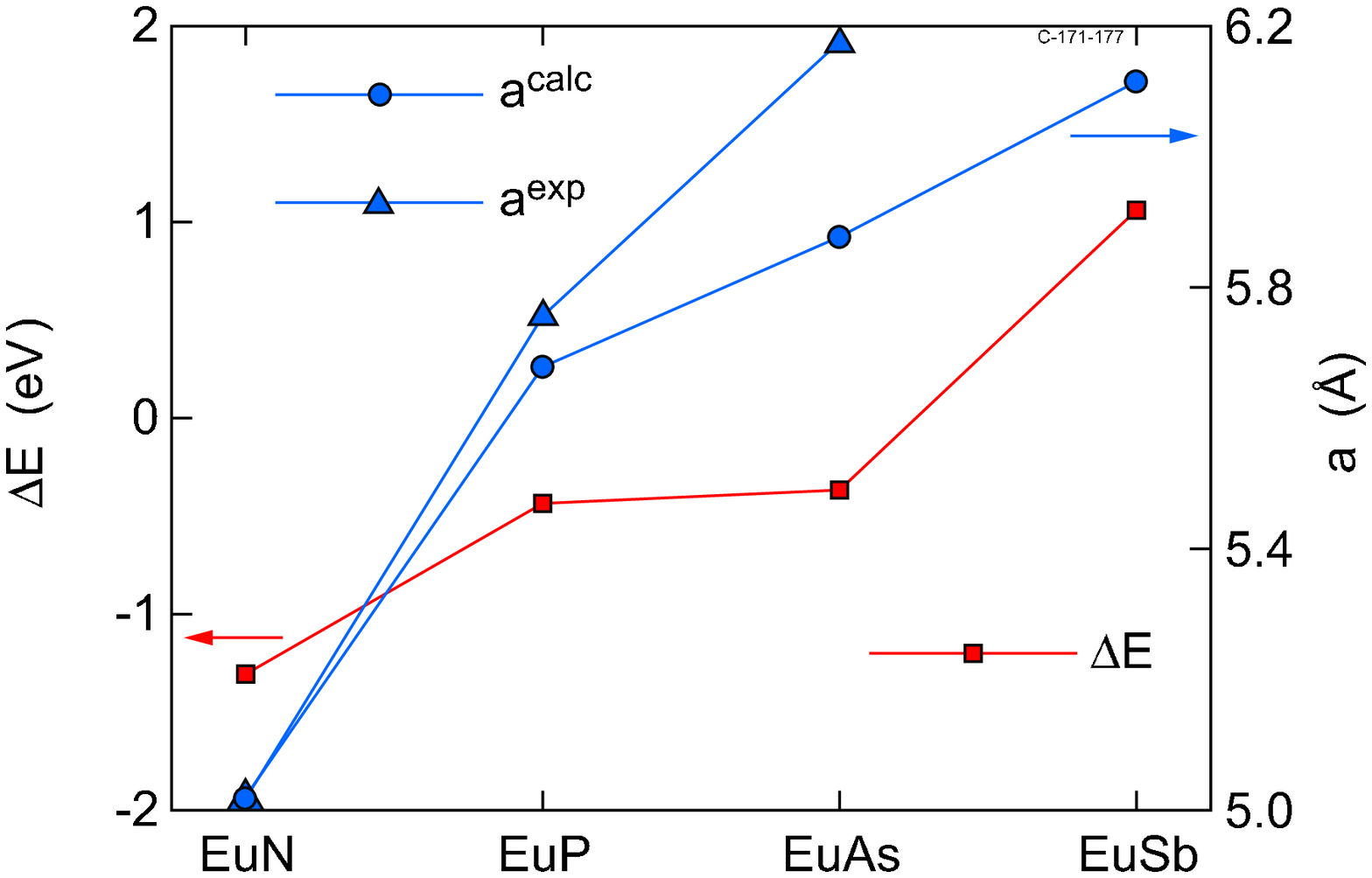}\\
\caption{The values of the lattice constant for the europium pnictides
from the SIC-LSD calculation. 
The circles represent the calculated values\cite{EuX} and 
the triangles are the experimental values.\cite{Villars}
Also shown
is the calculated energy difference between the divalent and trivalent form.
Negative values mean trivalent, positive values mean divalent.
\label{fig_EuPnict}
}
\end{center}
\end{figure}

Fig. \ref{fig_EuPnict} shows the equilibrium lattice constants for the europium
pnictides. Also shown are the energy differences between the divalent and
trivalent configurations. 
It is clear that EuN, EuP and EuAs are trivalent and EuSb is divalent.
These results are in full agreement with 
Hulliger\cite{hull} who states that Eu ions in the EuN and EuP are known to be
trivalent, while EuAs is known to contain some divalent ions. 
We are
not aware of any definitive measurement of the valence state of EuSb,
but clearly if the trend continues it will be divalent. 
Although the calculation predicts that EuAs, 
with the rocksalt crystal structure, is trivalent,
EuAs has in reality the Na$_2$O$_2$ crystal structure. 
This is a distortion of the NiAs
structure due to the formation of anion-anion pairs.\cite{hull}. 
We speculate that the occurence of Na$_2$O$_2$ crystal structure
and divalent ions is intimately connected.
These divalent ions do not occur in the rocksalt crystal structure
and hence explain the theoretical underestimate in the value of the 
lattice constants as seen in Fig. \ref{fig_EuPnict}.
The presence of divalent ions in EuAs would also make the valence energy 
difference curve of Fig. \ref{fig_EuPnict} more continuous.
These results
are qualitatively similar to those obtained for the ytterbium
pnictides\cite{yb1,yb2} (see below) where there is also an increasing tendency for
divalency as we go down the pnictide column of the Periodic Table.
However, in that case the divalent state is not reached.

\subsection{Yb monopnictides and monochalcogenides}

\begin{figure}
\begin{center}
\includegraphics[width=90mm,clip]{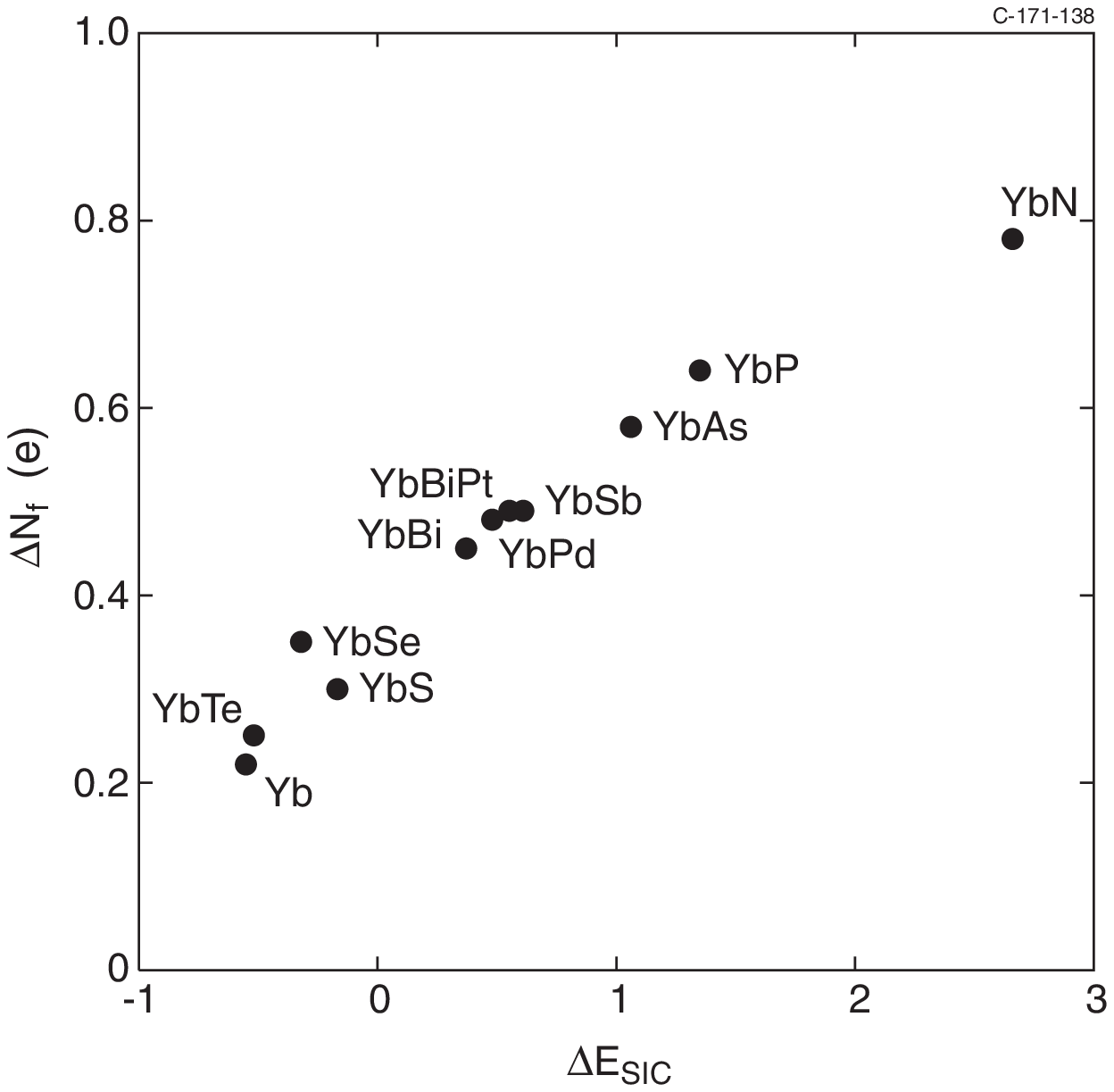}\\
\caption{
Divalent-trivalent energy difference (in eV) of Yb compounds
versus the total $f$-occupancy difference in the divalent and
trivalent configurations, as given by the SIC-LSD approach.
\label{fig_Yb_comp}}
\end{center}
\end{figure}

Here the application of the SIC-LSD method to Yb and a number
of its compounds is discussed.\cite{yb1,yb2} We concentrate in particular on the valence of Yb
ion in these systems, and some of the valence transitions, in particular
in YbS. Fig. \ref{fig_Yb_comp} shows for Yb compounds the $f$ electron
difference between the divalent and trivalent configurations versus their
energy difference. The behaviour is close to linear. The closest we come
to change of one electron occurs in YbN, specifically a change of 0.8 electrons,
for which also the highest energy difference between the two electronic
configurations is obtained. This maximum electron difference of 0.8 gradually decreases
to approximately 0.5 for YbPd, YbSb, YbBi and YbBiPt and for the four divalent Yb systems we
obtain between 0.22 to 0.35 electrons. For these divalent systems the difference in $f$
electron count has nearly disappeared and this seems to herald the arrival of
the divalent behaviour.

\begin{figure}
\begin{center}
\includegraphics[width=90mm,clip]{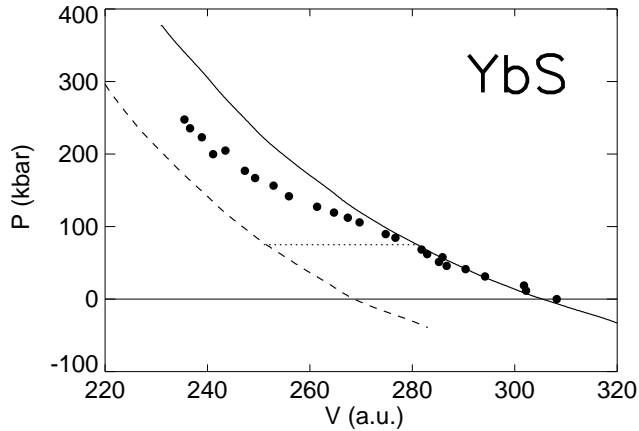}\\
\caption{
Equation of state for YbS as calculated by the SIC-LSD method.
The two theoretical curves correspond to 14 (solid line) and 13 (dashed line)
localized $f$ electrons, respectively, while the dots are the
experimental data of Ref. \onlinecite{syassen}. The dotted line marks the theoretical transition.
\label{fig_YbS}}
\end{center}
\end{figure}

Our calculations can also make contact with the pressure-volume measurements
which e.g. for YbS indicate anomalous behaviour around 100 kbar which could be
associated with intermediate valence\cite{syassen}. In particular, for this
system a trivalent state cannot seemingly be realized. In this case, our
study could shed some light on properties of this intermediate valence state.
From the common tangent construction of the total energies as a function of
volume for both divalent and trivalent YbS, we obtain with the SIC approach
a transition pressure which is $\sim $75 kbar and agrees well with the
experimental value of about $\sim $100 kbar (Fig. \ref{fig_YbS}). 
As seen in Fig. \ref{fig_YbS}, the experimental anomaly is not 
really correlated with an integer change in valence, but is as a matter of fact due to an $f$
electron delocalization. 
The quantitative theoretical description of this seemingly continuous
valence transition calls for a more elaborate theory than presented here.
The small change in $f$ electron occupancy of 0.3 electrons found with the SIC
approach upon delocalization  of an $f$ electron suggests the occurence of intermediate
valence of 2.3. This is consistent with the experimental estimate
of an intermediate valence of 2.4.\cite{syassen}

\section{Valence transitions in Actinide compounds}

\subsection{Elemental actindides}

Compared to the rare earth 4$f$-states, the 5$f$-states in the actinides are less inert, and in 
Th, Pa, and U, play an active role in
the cohesion, as manifested by the low symmetry crystal structures, low specific volumes, and
large bulk moduli. A localization transition occurs when going from
Pu to Am,~\cite{johansson} and in the later actinides, from Am to Es, the $f$-electrons are
non-bonding, high-symmetry crystal structures are attained, the specific volumes are large and
the bulk modulii relatively small. Pu is situated at the borderline between these two competing
pictures, and its very complex phase diagram implies that the $f$-electron properties are of
particularly intricate nature.
The low-symmetry
$\alpha $-Pu ground state is well reproduced by LDA calulations,~\cite{jones} whilst
the large-volume fcc phase of $\delta $-Pu, is believed to be characterized by localized $f$-electrons.

The SIC-LSD approach was applied to the calculation of the total energies as a function of atomic
volumes for the actinide elements from U to Fm.\cite{SSC,Act_to_be}
Ferromagnetic and paramagnetic arrangements were considered in the fcc structure, while 
antiferromagnetism was investigated in the hcp structure.
The overall total energy minima were found to
occur for the trivalent configuration in the case of Am to Cf, while Es and Fm have their minima
for the divalent state. 
Experimentally, Es is believed to be divalent, due to its large lattice constant,
while Fm has never been prepared in the solid state.
The U and Np metals have the lowest energies when the $f$-electrons are fully delocalized.
The position of Pu at the crossover between localized and delocalized $f$-electron behaviour is 
revealed in the very different outcome of the calculations depending on the magnetism. 
In a magnetic calculation, the trivalent configuration has
the lowest energy, however with a volume $\sim 30 \%$ higher than the experimental volume of $\delta$-Pu,
while a paramagnetic treatment leads to virtual degeneracy of any of the localization scenarios $f^0$
to $f^4$.
The trends of the energy difference between the divalent and trivalent configurations are qualitatively
the same in the actinides and the lanthanides and are governed by the relative stability of the half-filled
$f^7$ shell. 
In Table \ref{act_elements}, the calculated equilibrium volumes, bulk modulii, and transition
pressures for onset of delocalization are shown. Overall, the experimental data are well reproduced
by the SIC-LSD calculations.

\begin{figure}[t]
\begin{center}
\includegraphics[width=90mm,clip]{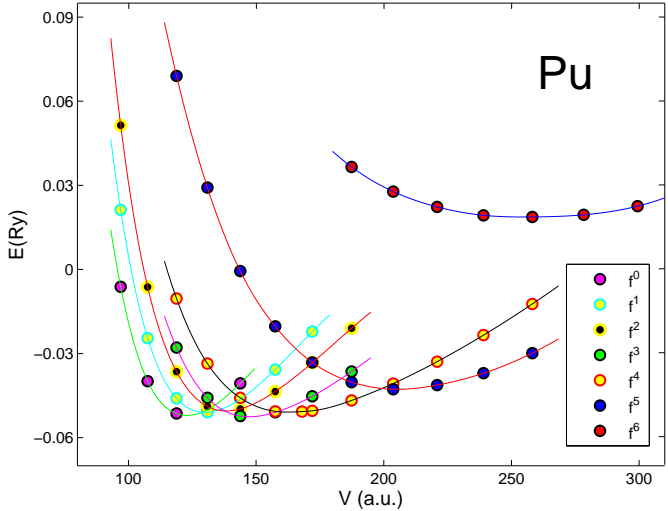}\\
\caption{Total energy of Pu in the paramagnetic fcc state.
        }
\label{pu}
\end{center}
\end{figure}

The peculiar role played by Pu in the series of elemental actinides is illustrated in  
figure \ref{pu}, which shows the total energy as a function of volume for all localization scenarios
from $f^0$ to $f^6$ in a paramagnetic treatment. Remarkably, within 0.03 eV/atom
the scenarios with 0,1,2,3 or 4 localized
$f$-electrons are degenerate, while
localizing 5 or  6 $f$-electrons is less favourable. The interpretation of this is that Pu at zero
pressure exists in a
complex quantum state vividly fluctuating between localized and delocalized $f$-electrons.
This is in accord with the experimental observation of several allotropic forms of Pu, including the
low-volume $\alpha$-phase and the high-volume $\delta$-phase, the latter being only stable at
elevated temperature or by alloying. In Table \ref{act_elements} we compare the data of the 
$f^4$ state to the $\delta$ phase.
If magnetic order is imposed,
in either ferro- or antiferro-magnetic arrangement, the $f^5$ configuration becomes the ground
state, i.e. formation of an ordered magnetic moment favours localization. The too large a
 volume associated with this solution indicates that this localization tendency is in fact 
overestimated,
as the SIC-LSD formalism does not account
properly for the strong quantum fluctuations taking place for a Pu ion in the metallic phase.
Experimentally, there is no sign of magnetic moments in elemental Pu.\cite{lashley}

\begin{figure}[t]
\begin{center}
\includegraphics[width=90mm,clip]{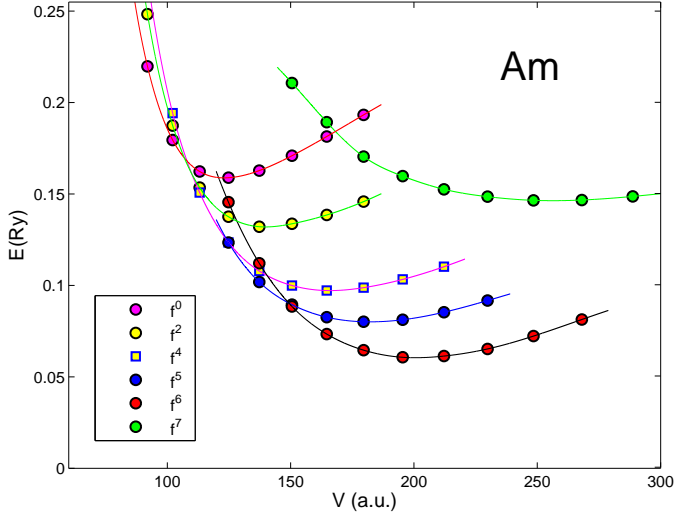}\\
\caption{Total energy versus volume for Americium in the paramagnetic fcc structure.
Several configurations of the localized $5f$ subshell are considered. 
        }
\label{am}
\end{center}
\end{figure}

In Fig. \ref{am} the total energy as a function of volume is shown
for several $f^{n}$ configurations in Am.
As already mentioned, the trivalent $f^{6}$ configuration has the lowest energy,
with an equilibrium volume of 201 a.u., while the experimental
volume is 198 a.u.. 
Upon compression the band
formation energy increases and the energy difference between $f^{5}$ and $%
f^{6}$ decreases. At a volume of about 140 a.u. the $f^{5}$ and $f^{6}$
total energy curves cross, and from then on the $f$-electron band states
start to dominate. Around $V=100$ a.u. the $f$-delocalization is complete. 
This is in good agreement with the experimental
observation of a structural phase transition (AmII-AmIII) taking place at 100 kbar and at
a compression corresponding to
$77 \%$ of the equilibrium volume,\cite{heathman} which is
interpreted as the onset of $f$-bonding in Am. A rough theoretical estimate
of the transition pressure, given by the slope of the $E(V)$ curves at the
crossing point, leads to  a value of 160 kbar.
This estimate constitutes an upper bound to the delocalization pressure, since
the high-valence SIC-LSD configurations represent a rather limited many-body
wavefunction for a correlated state with actively bonding $f-$electrons. In
addition, the correct (orthorhombic) high pressure phase was approximated here by
the fcc structure, and finite temperature may also lower the phase transition
pressure. Experimentally, the occurrence of the AmIV phase at $V/V_0=0.63$ at $p=175$ 
kbar probably marks the endpoint of the $f$-electron delocalization process in Am.

\begin{figure}
\begin{center}
\includegraphics[width=90mm,clip]{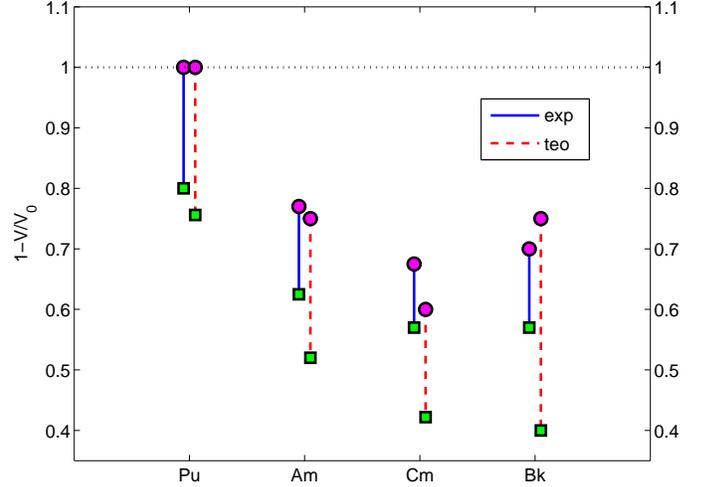}\\
\caption{
\label{deloc}
Trends in the ranges of volumes of $f$ electron delocalization in the actinide elements.
Onset (completion) of delocalization is marked with balls (squares), for each element
with experimental data (blue and full line)
to the left and theoretical data (red and dashed line) to the right.
Volumes are given relative to the (localized) equilibrium volume at ambient conditions. 
Experimental ranges are defined by the smallest
volumes observed in the high-symmetry (fcc) phase and the largest volumes observed for the
low-symmetry ($\alpha$-U type) phases in high pressure experiments
(Refs. \onlinecite{heathman,landerCm,benedict}; for Pu the range is defined by the zero pressure
volumes of the $\delta$- and $\alpha$-phases).
The theoretical ranges are calculated within the fcc structure only.
}
\end{center}
\end{figure}

Figure \ref{deloc} shows
the upper bounds for a transition pressure to a phase
with bonding $f-$electrons. The agreement with experiment varies accross the
series, but is generally not bad (within a factor of 2).
This qualitative agreement
shows that the magnitude of the SIC-LSD localization energy is roughly correct.
Figure \ref{deloc} also shows the comparison of the volume ranges over which
$f$-electron delocalization occurs for the actinide elements Pu, Am, Cm and Bk, as given by this theory and
experiment. The qualitative trend as given by the relative stability of Cm($f^7$), is well
reproduced by the calculations, with a tendency of the calculated end-points of $f$-transition to be 
at too low a volume.

\subsection{Actinide monopnictides and monochalcogenides}

\begin{figure}
\begin{center}
\includegraphics[width=90mm,clip]{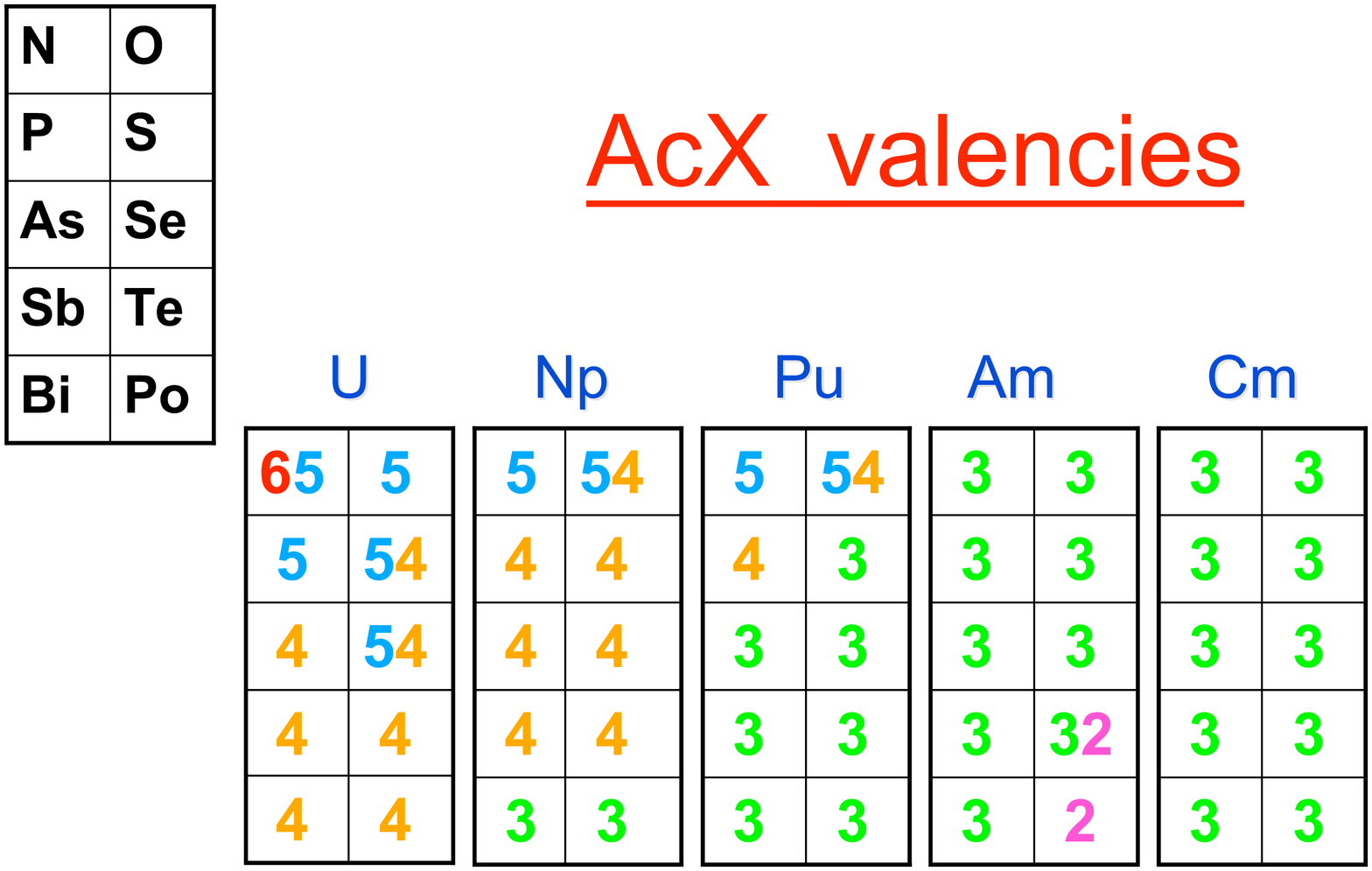}\\
\caption{Trends in localization through the AcX series. For each actinide, Ac=U, Np, Pu, Am, Cm, a block of
10 ligands are considered; the pnictides X=N, P, As, Sb, Bi, and the chalcogenides X= O, S, Se, Te, and Po.
The numbers designate the calculated Ac valence for that particular AcX compound. Where two numbers are given,
the corresponding valences are degenerate.}
\label{chalcpnic}
\end{center}
\end{figure}

In the actinide monopnictides and monochalcogenides, which all crystallize in the NaCl structure
at ambient conditions, the actinide-actinide separations are larger than in the elemental metals, and the
tendency towards $f$-electron localization can already be observed from Np compounds onwards.
Figure \ref{chalcpnic} displays the calculated SIC-LSD ground state configurations through the series of 
U-, Np-, Pu-, Am-,
and Cm-mono-pnictides and -mono-chalcogenides. The calculations reveal clear trends towards more and more
actively bonding $f$-electrons for a) lighter actinides, and b) lighter ligands. For the lighter actinides,
the $f$-orbitals are more extended, leading to larger overlaps with their nearest neighbours, and smaller
self-interaction corrections, both of these effects favouring band formation. For the lighter ligands, in
particular N and O, both the volume is decreased and ionicity is larger, the first of these  effects
leading to larger direct actinide-actinide overlap, and the latter effect favouring charge transfer.

The Cm compounds are the most localized systems, all exhibiting Cm in the trivalent $f^7$ configuration.
The $f^7$ shell is so stable that variations of the ligand cannot disrupt its stability, and scenarios 
with either one more or less localized $f$-electron have distinctly higher energies. Trivalency 
prevails in the Am-compounds, but the stability of the $f^7$ shell causes the divalent Am state to be 
important in AmTe and AmPo. In the Pu
compounds, the trivalent state also dominates, but for the lighter ligands $f$-electron delocalization 
sets in. In the Np compounds the tetravalent state dominates, while in the U compounds pentavalent 
states occur for the lighter
ligands.

\begin{figure*}[t]
\begin{tabular}{cc}
\includegraphics[width=70mm]{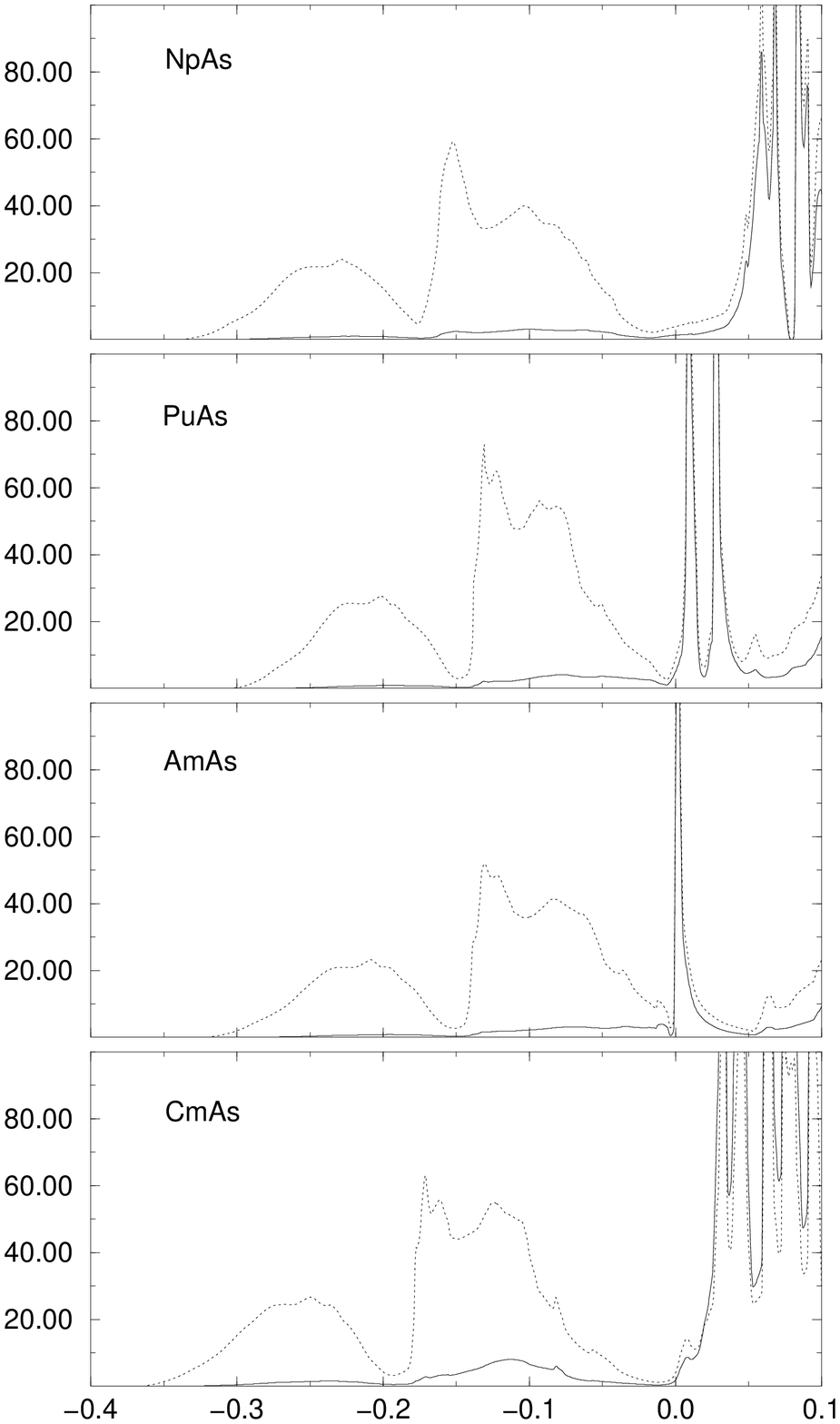}&
\includegraphics[width=70mm]{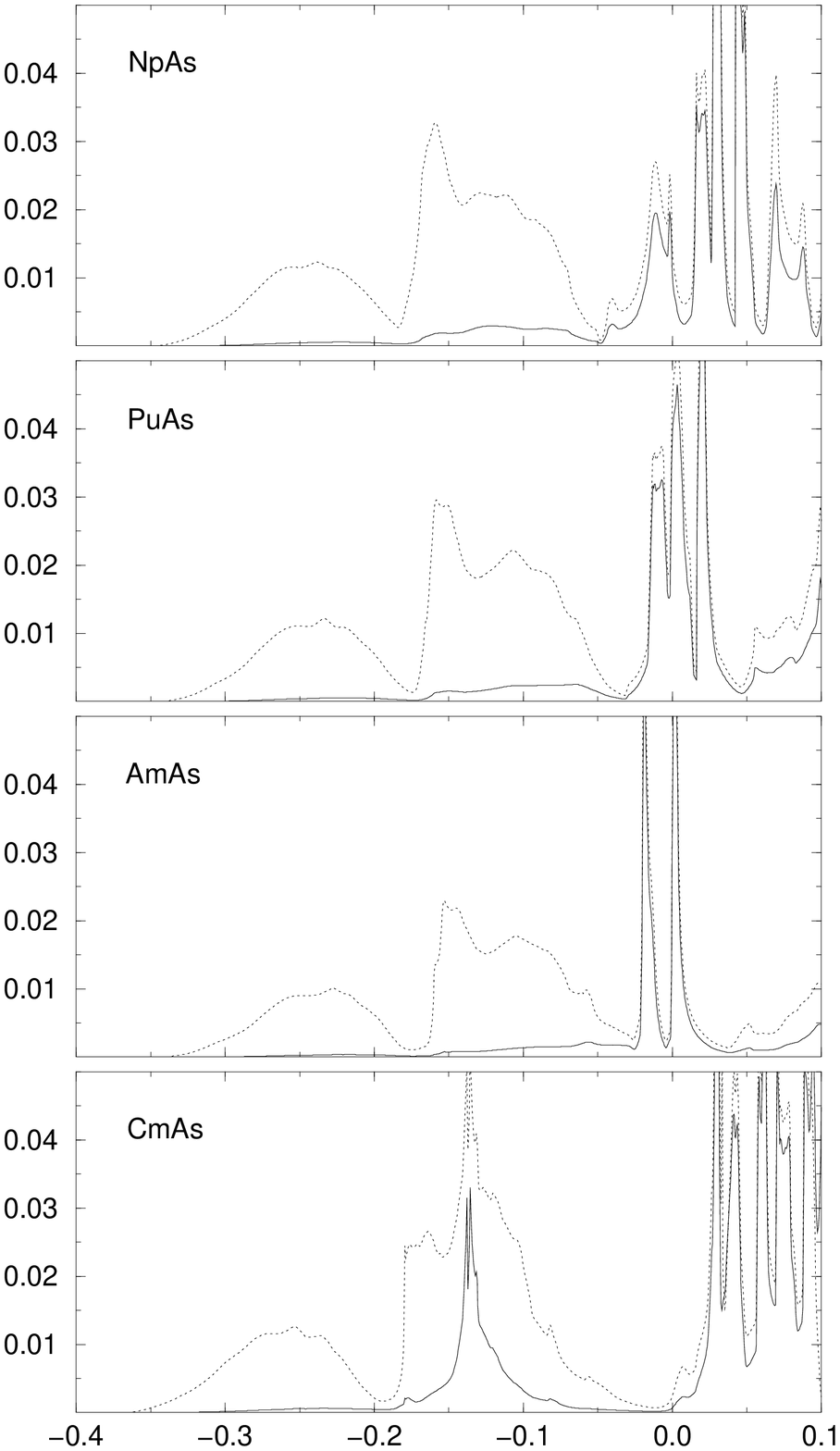} 
\end{tabular}
\caption{Densities of states (in states/(Ry formula unit)) 
for the actinide arsenides NpAs, PuAs, AmAs, CmAs, 
with  a trivalent (left), or tetravalent (right) actinide ion, 
respectively. The solid and dotted lines represent the $f$ projected- and  total 
densities of states, respectively. The energies are given in Ry, with the Fermi level
at energy zero.
\label{DOS}
}

\end{figure*}

The density of states (DOS) of the actinide arsenides are shown in Fig. \ref{DOS}, 
with both trivalent and tetravalent actinide ions. 
In the trivalent case, the non-(SIC) $f$-degrees of freedom give rise to narrow unoccupied bands above the Fermi
level. In the tetravalent case the additional delocalized $f$-electron appears as an extra $f$-band. In Cm, this
band appears far below the Fermi level, while in Am, Pu and Np this band lies just below the Fermi level.
The band formation due to this extra band is sufficiently large in NpAs to outweigh the localization energy,
and the tetravalent configuration becomes the ground state.

\subsection{Uranium compounds UPt$_3$ and UPd$_3$}

\begin{figure}
\begin{center}
\includegraphics[width=90mm,clip]{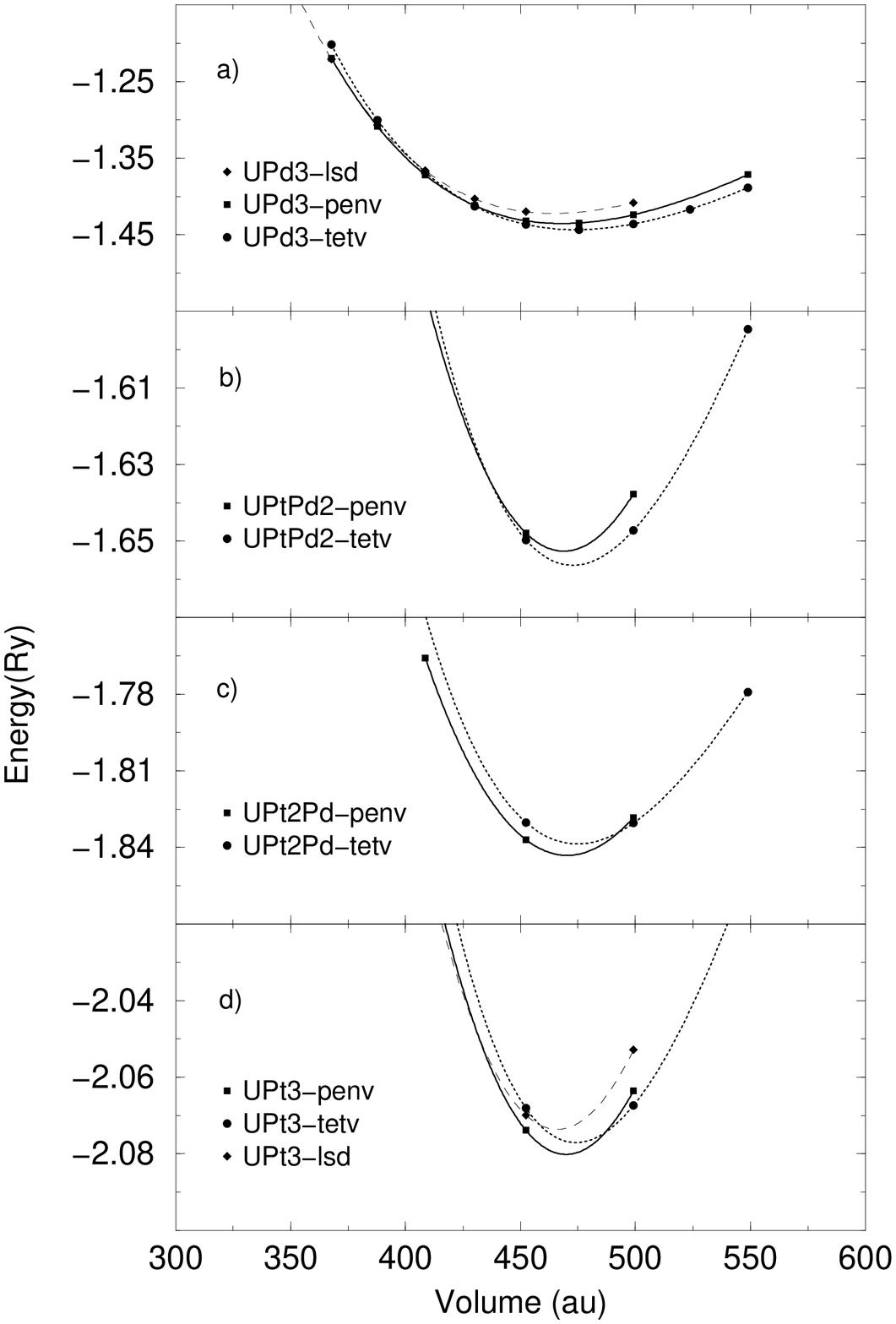}\\
\caption{Total Energy (in Ry/f.u.) versus volume (in (a.u.)$^{3}$) for a) UPd$_3$,
b) UPtPd$_2$, c) UPt$_2$Pd, and d) UPt$_3$.
}
\label{updpt}
\end{center}
\end{figure}

Using the SIC-LSD method, we have similarly studied the 5 $f$ valence configurations of U in
UPt$_3$ and UPd$_3$.~\cite{prlupd3} Both compounds are isoelectronic as far as the valence electrons are concerned, 
however the 5$d$ electrons are less tightly bound to the nucleus than the 4$d$ electrons, which
results in a increased overlap of the U $f$ and the transition metal $d$ orbitals in UPt$_3$. 
Correspondingly, the total energy calculations determine the tetravalent $f^2$ and the pentavalent
$f^1$ configurations
for UPd$_3$ (Fig. \ref{updpt}a) and UPt$_3$ (Fig. \ref{updpt}d) respectively.
The calculated equilibrium volumes, $V_{UPd_3}=474.1($a.u.$^3$) and $V_{UPt_3}=469.7($a.u.$^3$),  
compare well with the experimental values of
$V_{UPd_3}^{exp.}=469.5($a.u.$^3$) and $V_{UPt_3}^{exp}=472.9($a.u.$^3$).~\cite{villars}
In both UPd$_3$ and UPt$_3$, the small energy differences between the
tetravalent and pentavalent configurations
indicate near degeneracy (within $\sim$5 mRy), rather
than a clearly preferred $f^2$ groundstate for UPd$_3$ and an $f^1$ groundstate for UPt$_3$.
This in a way mirrors the results from XPS measurements, where in UPt$_3$ a weak shoulder
at the Fermi energy indicates the presence of itinerant 5$f$ electrons,~\cite{schneider}
and in UPd$_3$ this same shoulder
is seen about 1 eV below the Fermi level, and is interpreted in terms of localized $f$
electrons.~\cite{baer}  Sharp features, usually associated with localized $f$-electrons,
are not observed.
In UPt$_{3-x}$Pd$_x$
alloys, gradually substituting Pt by Pd, 
leads to a transition from the
pentavalent to tetravalent groundstate between UPt$_2$Pd and UPtPd$_2$, as can be seen from
Figs. \ref{updpt}c and \ref{updpt}b respectively.
An additional $f$-electron gets localized, a transition that appears to be driven 
mainly by the changes
in the electronic structure related to the decreasing $fd$ hybridization when replacing
Pt by Pd.
The
opposite effect is obtained by putting UPd$_3$ under pressure. The total energy
calculations for UPd$_3$  show that the pentavalent configuration
becomes energetically favourable 
at pressures of approximately 250 kbar.
From the observed similarities in the pentavalent DOS of both UPd$_3$ and UPt$_3$~\cite{prlupd3},
and given that UPt$_3$ is a heavy fermion material at zero pressure, we expect 
UPd$_3$ to become heavy fermion under pressure. 
Experimental efforts\cite{xxx} to see the proposed transtion of UPd$_3$ under pressure failed. 
This can be either due to an underestimate in the present theory of the
transition pressure, or finite temperature effects in the experiment.

\section{Valence transition in Ce at finite temperature}
 
Cerium is the first element in the Periodic Table that contains an
$f$ electron, and shows an
interesting phase diagram.\cite{KoskenmakiGschneidner:78} In
particular, the isostructural (fcc $\to$ fcc) $\alpha-\gamma$ phase
transition is associated with a 15\%-17\% volume collapse and
total quenching of the magnetic moment.\cite{KoskenmakiGschneidner:78} 
The low-pressure $\gamma$-phase shows a local magnetic moment, and is
associated with a trivalent configuration of Ce ion. At the temperatures
in which the $\gamma$-phase is accessible, it is in a paramagnetic
disordered local moment state.  Increasing the pressure, the material
first transforms into the $\alpha$-phase, which is indicated to be in an
intermediate valence state with quenched magnetic moment. At high
pressures (50 kbar at room temperature) Ce eventually transforms into
the tetravalent $\alpha'$-phase.  With increasing temperature, the
$\alpha-\gamma$ phase transition shifts to higher pressures, ending in
a critical point (600K, 20 kbar), above which there is a continuous
crossover between the two phases.

To describe the full phase diagram of the Ce $\alpha-\gamma$ phase
transition we have modelled Ce as a pseudoalloy,\cite{alloy_borje,alloy_axel}
in the spirit of the Hubbard III approximation \cite{Hubbard3},
consisting of the trivalent (SIC-LSD) Ce atoms with concentration c,
and the tetravalent (LDA) Ce atoms with the concentration (1-c).  In
order to properly take into account the disordered local moments of the
trivalent Ce atoms in the $\gamma$-phase, equal probabilities for up
and down orientations of their local moments were adopted.
Assuming homogeneous randomness, this ternary
pseudoalloy can be described by the coherent potential approximation
(CPA). The respective concentrations of the trivalent and the
tetravalent Ce ions in the pseudoalloy are then determined by
minimizing the free energy for each volume and temperature with 
respect to the concentration $c$.

\begin{figure} 
\begin{picture}(0,0)%
\includegraphics{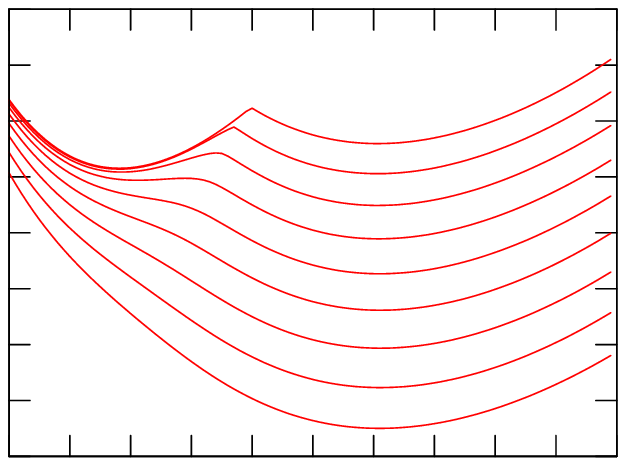}%
\end{picture}%
\begingroup
\setlength{\unitlength}{0.0200bp}%
\begin{picture}(12599,8640)(0,0)%
\put(2750,1650){\makebox(0,0)[r]{\strut{}-0.7}}%
\put(2750,2455){\makebox(0,0)[r]{\strut{}-0.695}}%
\put(2750,3260){\makebox(0,0)[r]{\strut{}-0.69}}%
\put(2750,4065){\makebox(0,0)[r]{\strut{}-0.685}}%
\put(2750,4870){\makebox(0,0)[r]{\strut{}-0.68}}%
\put(2750,5675){\makebox(0,0)[r]{\strut{}-0.675}}%
\put(2750,6480){\makebox(0,0)[r]{\strut{}-0.67}}%
\put(2750,7285){\makebox(0,0)[r]{\strut{}-0.665}}%
\put(2750,8090){\makebox(0,0)[r]{\strut{}-0.66}}%
\put(3025,1100){\makebox(0,0){\strut{} 140}}%
\put(3900,1100){\makebox(0,0){\strut{} 150}}%
\put(4775,1100){\makebox(0,0){\strut{} 160}}%
\put(5650,1100){\makebox(0,0){\strut{} 170}}%
\put(6525,1100){\makebox(0,0){\strut{} 180}}%
\put(7400,1100){\makebox(0,0){\strut{} 190}}%
\put(8275,1100){\makebox(0,0){\strut{} 200}}%
\put(9150,1100){\makebox(0,0){\strut{} 210}}%
\put(10025,1100){\makebox(0,0){\strut{} 220}}%
\put(10900,1100){\makebox(0,0){\strut{} 230}}%
\put(11775,1100){\makebox(0,0){\strut{} 240}}%
\put(550,4870){\rotatebox{90}{\makebox(0,0){\strut{}$F(T,V)$ [Ry]}}}%
\put(7400,275){\makebox(0,0){\strut{}$V$ [a.u.$^3$]}}%
\end{picture}%
\endgroup
\caption{\label{fig:Efree-eq} 
The free energies as function of the volume for the temperatures
0 (highest curve), 200, 400, 600, 800, 1000, 1200, 1400 and 1600 K (lowest curve).
The zero of energy is arbitrary.
}
\end{figure}

The free energy of the physical system at a given volume
can be obtained
by evaluating the concentration dependent free energy at the
minimizing  concentration $c_{\rm min}$:
\begin{equation}
F(T,V) = F(T,c_{\rm min},V) \,.
\end{equation}
These free energies are displayed in Fig.~\ref{fig:Efree-eq}, which clearly shows a
double-well behaviour, corresponding to the two separate phases of Ce, at low temperatures, 
which is gradually smoothened out with increasing
temperatures. Furthermore one finds that, at elevated temperatures, the free energy is mainly
lowered for large lattice constants, corresponding to the $\gamma$-phase, with its larger entropy.
\unitlength1cm
\begin{figure}[b]
\begin{picture}(10,5)
\put(0.1,2.9){\rotatebox[origin=center]{90}{$p$ [kbar]}}
\put(3.8,0){$V$ [a.u.$^{3}$]}
\put(0.5,0.15){\includegraphics[scale=0.75]{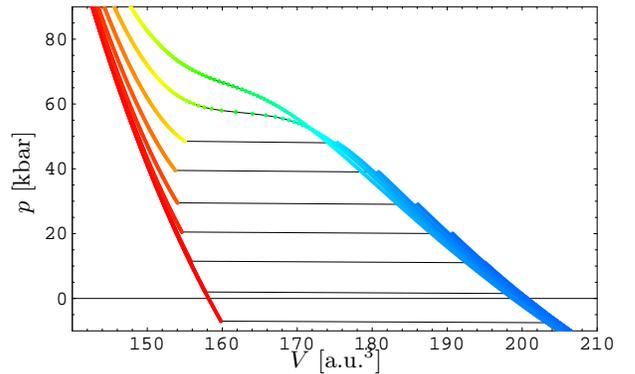}}
\end{picture}
\caption{\label{Ce-p-V} 
  Calculated isotherms for the temperatures
  T=0 (lowest curve), 200, 400, 600, 800, 1000, 1200, 1400, and, 1600K (highest curve).  The color
  indicates the fraction of localized electrons: blue is all localized
  ($\gamma$-phase) and red is all delocalized ($\alpha$-phase).}
\end{figure}
Inserting the minimizing concentration $c_{\rm min}$ into the pressure-volume
relation
\begin{equation}
\label{eq:p-V}
p(T,V) = p(T,c_{\rm min},V) = -\frac{\partial}{\partial V} F(T,c_{\rm min},V),
\end{equation}
allows to calculate the isotherms of Ce, which are displayed in
Fig. \ref{Ce-p-V}. It can be seen that the average valence, close to the
coexistence line, gradually changes with increasing temperature. Above the
critical temperature, the valence changes continuously with increasing pressure
from trivalent to tetravalent Ce ion.

\begin{figure}[h] 
\hspace{-7.0cm}
\begin{minipage}{7cm}
\vspace*{-1.2cm}
\begin{picture}(0,0)%
\includegraphics{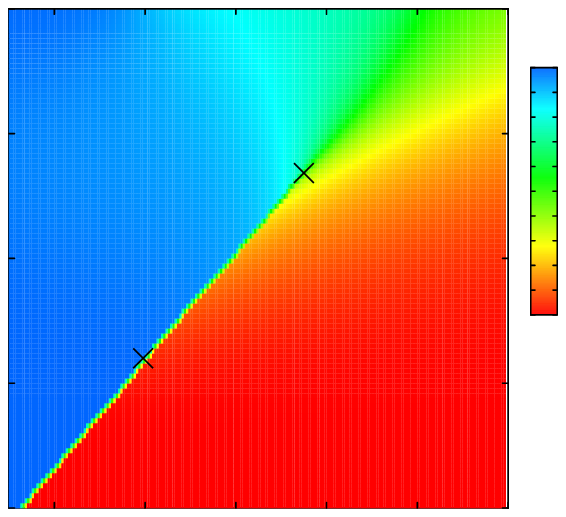}%
\end{picture}%
\begingroup
\setlength{\unitlength}{0.0200bp}%
\begin{picture}(19800,11880)(0,0)%
\put(14477,5402){\makebox(0,0)[l]{\strut{} 0}}%
\put(14477,5758){\makebox(0,0)[l]{\strut{} 0.1}}%
\put(14477,6114){\makebox(0,0)[l]{\strut{} 0.2}}%
\put(14477,6470){\makebox(0,0)[l]{\strut{} 0.3}}%
\put(14477,6826){\makebox(0,0)[l]{\strut{} 0.4}}%
\put(14477,7182){\makebox(0,0)[l]{\strut{} 0.5}}%
\put(14477,7538){\makebox(0,0)[l]{\strut{} 0.6}}%
\put(14477,7894){\makebox(0,0)[l]{\strut{} 0.7}}%
\put(14477,8250){\makebox(0,0)[l]{\strut{} 0.8}}%
\put(14477,8606){\makebox(0,0)[l]{\strut{} 0.9}}%
\put(14477,8962){\makebox(0,0)[l]{\strut{} 1}}%
\put(9900,1081){\makebox(0,0){\strut{}$p$ [kbar]}}%
\put(4437,6215){\rotatebox{90}{\makebox(0,0){\strut{}$T$ [K]}}}%
\put(6960,1906){\makebox(0,0){\strut{} 0}}%
\put(8267,1906){\makebox(0,0){\strut{} 20}}%
\put(9574,1906){\makebox(0,0){\strut{} 40}}%
\put(10880,1906){\makebox(0,0){\strut{} 60}}%
\put(12187,1906){\makebox(0,0){\strut{} 80}}%
\put(13494,1906){\makebox(0,0){\strut{} 100}}%
\put(5949,2621){\makebox(0,0)[r]{\strut{} 0}}%
\put(5949,4418){\makebox(0,0)[r]{\strut{} 500}}%
\put(5949,6215){\makebox(0,0)[r]{\strut{} 1000}}%
\put(5949,8012){\makebox(0,0)[r]{\strut{} 1500}}%
\put(5949,9809){\makebox(0,0)[r]{\strut{} 2000}}%
\end{picture}%
\endgroup
\end{minipage}
\caption{ Phase diagram obtained for the pseudoalloy, composed of $\alpha$-
  and $\gamma$-Ce.  The crosses indicate the calculated and
  experimental critical points.
\label{full-phase-diag}}
\end{figure}

In Fig. \ref{full-phase-diag} we present the phase diagram, obtained
from the free energies of the $\alpha$-$\gamma$ pseudoalloy, with
the $\gamma$-phase described by the DLM approach. It can clearly be seen
in the figure how the transition becomes continuous above the
critical temperature. The experimentally observed critical point (600K, 20 kbar)
falls on top of the calculated phase separation line, which starts at the zero
temperature transition pressure of -7.4 kbar. This means that the slope of the
phase separation line is in very good agreement with experiments. The calculated critical
temperature overestimates the experimental one by roughly a factor of 2,
which is still reasonable considering that the critical temperature
is very sensitive to various small details of the calculations and in
particular the theoretical lattice parameters of both the Ce phases.
Note that our calculated value of 169 K for the $T_c$ at zero pressure compares
well with the experimental value of 141$\pm$10 K.

\begin{figure}
\begin{picture}(0,0)%
\includegraphics{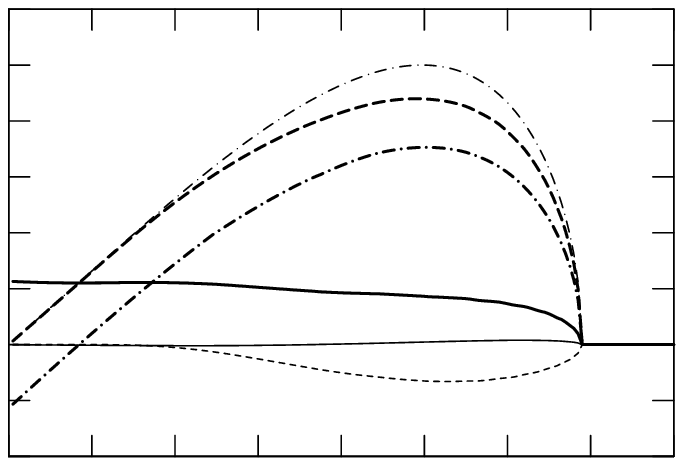}%
\end{picture}%
\begingroup
\setlength{\unitlength}{0.0200bp}%
\begin{picture}(12599,8640)(0,0)%
\put(1925,1650){\makebox(0,0)[r]{\strut{}-4}}%
\put(1925,2455){\makebox(0,0)[r]{\strut{}-2}}%
\put(1925,3260){\makebox(0,0)[r]{\strut{} 0}}%
\put(1925,4065){\makebox(0,0)[r]{\strut{} 2}}%
\put(1925,4870){\makebox(0,0)[r]{\strut{} 4}}%
\put(1925,5675){\makebox(0,0)[r]{\strut{} 6}}%
\put(1925,6480){\makebox(0,0)[r]{\strut{} 8}}%
\put(1925,7285){\makebox(0,0)[r]{\strut{} 10}}%
\put(1925,8090){\makebox(0,0)[r]{\strut{} 12}}%
\put(2200,1100){\makebox(0,0){\strut{} 0}}%
\put(3397,1100){\makebox(0,0){\strut{} 200}}%
\put(4594,1100){\makebox(0,0){\strut{} 400}}%
\put(5791,1100){\makebox(0,0){\strut{} 600}}%
\put(6988,1100){\makebox(0,0){\strut{} 800}}%
\put(8184,1100){\makebox(0,0){\strut{} 1000}}%
\put(9381,1100){\makebox(0,0){\strut{} 1200}}%
\put(10578,1100){\makebox(0,0){\strut{} 1400}}%
\put(11775,1100){\makebox(0,0){\strut{} 1600}}%
\put(550,4870){\rotatebox{90}{\makebox(0,0){\strut{}$\Delta E$ [mRy]}}}%
\put(6987,275){\makebox(0,0){\strut{}$T$ [K]}}%
\end{picture}%
\endgroup
\caption{
\label{fig:jumps}
Discontinuities of the total energy (thick solid line), the total
entropy $T S$ (thick dashed line) and the $p V$ term (thick
dashed-dotted line) over the phase separation line as function of the
temperature. The entropy term is further decomposed into the
electronic (thin solid line), the mixing (thin dashed line) and the
magnetic (thin dashed-dotted line) contribution.}
\end{figure}

Finally we examine in more detail the discontinuity across the phase
separation line. Fig. \ref{fig:jumps} shows the magnitude of the
discontinuities for the various contributions of the Gibbs free energy.
As expected, all contributions vanish at the critical temperature,
above which there is a continuous cross-over between the $\alpha$- and
the $\gamma$-phase.  It also can be seen from this figure that the
entropy discontinuity is by far the largest contribution. The phase
transition is therefore driven by entropy, rather than by energetics.
The entropy discontinuity itself is mainly due to the magnetic
entropy. Thus it is the entropy, and not the internal energy that drives 
the Ce $\alpha-\gamma$ phase transition.
Analysis of experimental data leads to the same conclusion.\cite{amadon}


\section{Summary and Conclusion}

In this article a review has been given of some applications of the SIC-LSD to the calculation
of the valences of $f$-electron systems. A methodology has been presented which is able to determine valence changes
as a function of pressure and chemical composition. An important finding of this
work is the discovery of the dual character of $f$ electrons. The number of localized $f$-electrons
defines the valence. Furthermore, a correlation between a change in valence and
the number of band-like $f$-electrons has been established.

The versatility and functionality of the SIC-LSD has increased with the L-SIC. In particular as
the example of the p-T plot in Fig. \ref{full-phase-diag} shows a finite T generalization of the method has been
successfully developed. This will allow us to perform the finite T study of the p-V curves of several of
the compounds discussed in this paper and to determine if the continuous transitions as some times seen
experimentally can be obtained.

\section{acknowledgements}

This work was partially funded by the EU Research Training Network
(contract:HPRN-CT-2002-00295) 'Ab-initio Computation of Electronic 
Properties of $f$-electron Materials'. AS acknowledge support from the Danish Center for Scientific
Computing.  Work of LP was
sponsored by the Office of Basic Energy Sciences, U.S. Department of Energy.

\widetext

\begin{table}
\caption[b]{Calculated transition pressures for the electronic and structural
phase transitions in the cerium monopnictides and monochalcogenides. 
Also quoted are the specific volumes (relatively to the zero pressure equilibrium 
volume)
on the two sides of the transition  Ref. \onlinecite{errorbars}. The notation (d)
and
(l) refers to calculations with delocalized or localized Ce $f$-electrons, i.e.
tetravalent or trivalent Ce atoms. B2${}^*$ denotes the distorted B2 structure.\\
The subscripts refer to the following references: a: Ref. \onlinecite{Vedel};
b: Ref. \onlinecite{Mori};
c: Ref. \onlinecite{Werner};
d: Ref. \onlinecite{Leger-cesb};
e: Ref. \onlinecite{Leger-cebi};
f: Ref. \onlinecite{Leger};
g: Ref. \onlinecite{Croft};
h: Ref. \onlinecite{cese-exp};
i: Ref. \onlinecite{cete-exp};
j: Ref. \onlinecite{Shi2}. 

\label{Table1}
}
\bigskip
\begin{tabular}{|lc|cccccc|}
\hline
 compound    & transition  &
    \multicolumn{2}{c}{P$_t$ (kbar)} &
    \multicolumn{2}{c}{$V_1/V_0$ } &
    \multicolumn{2}{c|}{$V_2/V_0$ } \\
     &   & theo. & expt. & theo. & expt. & theo & expt. \\
\hline
CeN   &  B1(d) $\rightarrow$ B2(d)      &  620 &  -      & 0.760&  -  & 0.724 &
 -  \\
CeP   &  B1(l) $\rightarrow$ B1(d)      &   71 & 90$^a$,55$^b$   & 0.933 & 0.89$^a$
 &  0.853 & 0.85$^a$ \\
CeP   &  B1(d) $\rightarrow$ B2(d)      &  113 & 150(40)$^a$,250$^j$ & 0.827 & 0.82$^a$ &
0.706 & 0.71$^a$ \\
CeAs  &  B1(l) $\rightarrow$ B2(d)      & 114  & 140(20)$^c$,210$^j$ & 0.893 & 0.84$^c$ &
0.713 & 0.73$^c$ \\
CeSb  &  B1(l) $\rightarrow$ B2*(l)     &  70  & 85(25)$^d$,150$^j$  & 0.922 & 0.90$^d$ &
0.813 & 0.80$^d$ \\
CeSb  &  B2*(l) $\rightarrow$ B2*(d)    &  252 &  -      & 0.717 &  -  &  0.680  &
 -  \\
CeBi  &  B1(l) $\rightarrow$ B2*(l)     &   88 & 90(40)$^e$  & 0.897 & 0.87$^e$ &
0.789 & 0.78$^e$  \\
CeBi  &  B2*(l) $\rightarrow$ B2*(d)    &  370 &  -      & 0.666 &  -  &  0.638  &
 -  \\
CeS   &  B1(l) $\rightarrow$ B1(d)      &  101 &  -$^f$,125(15)$^g$   &
                          0.918& 0.93$^g$ &  0.855& 0.88$^g$ \\
CeS   &  B1(d) $\rightarrow$ B2(l)      &  243 &  -          &
                          0.788 &      -  & 0.742 &  -      \\
CeS   &  B2(l) $\rightarrow$ B2(d)      &  295 &  -          &
                          0.724 &      -  & 0.688 &  -     \\
CeSe  &  B1(l) $\rightarrow$ B2(l)      & 124  & 170(30)$^h$ &
                          0.890& 0.86$^h$& 0.779 & 0.77$^h$ \\
CeSe  &  B2(l) $\rightarrow$ B2(d)      & 377  & -           &
                          0.683 &      -  & 0.652 &  -      \\
CeTe  &  B1(l) $\rightarrow$ B2(l)     &  74  & 55(25)$^i$  &
                          0.915 & 0.93$^i$ & 0.798 & 0.84$^i$ \\
CeTe  &  B2(l) $\rightarrow$ B2(d)     &  435 &   -         &
                          0.647 &    -    & 0.623  &   -     \\
\hline
\end{tabular}
\newline
\bigskip

\end{table}

\begin{table}[htb]
\renewcommand{\arraystretch}{1.5}
\caption{Calculated isostructural  transition pressures, $P_t$ (in GPa; 1 GPa=10 kbar),
and volume changes (in \%),
of Sm monochalcogenides. Experimentally, the transition of SmS is
discontinuous, while those of SmSe and SmTe
(at room temperature) are continuous.\\
The subscripts refer to the following references: 
$^d$: Ref.      \onlinecite{Benedict1};
$^e$: Insulator-metal transition of Ref.      \onlinecite{Sidorov};
$^f$: Present author's estimates from figures of Ref.      \onlinecite{LeBihan} and
$^g$: Ref.      \onlinecite{tsiok}. The volume changes for SmSe and SmTe are obtained by
extrapolation over the transition range.
\label{Ptranssm}}
\bigskip
\begin{tabular} {|l|cc|cc|}
\hline
Compound & \multicolumn{2}{c|}{$P_t$(GPa)} & \multicolumn{2}{c|}{Volume collapse (\%)} \\

      & Theory & Expt.                     & Theory & Expt. \\

\hline
SmS   & 0.1  & $0.65^d$,             $1.24^e$           & 11.1  & $13.5^d$, 13.8$^e$ \\
SmSe  & 3.3  & $\sim 4^d$, $3.4^e$, $3-9^f$,$2.6-4^g$   & 9.8   & $8^d$, $ 11^f$, $7^g$ \\
SmTe  & 6.2  & $2-8^d$, $5.2^e$, $6-8^f$,   $4.6-7.5^g$ & 8.4   &        $9^f$,$7^g$  \\
\hline
\end{tabular}

\end{table}

\begin{table}[htb]
\renewcommand{\arraystretch}{1.5}
\caption{Calculated isostructural  transition pressures, $P_t$ (in GPa; 1 GPa=10 kbar),
and volume changes (in \%),
of Eu monochalcogenides.\\
The subscripts refer to the following references:
$^a$: Ref.      \onlinecite{Jayaraman3};
$^b$: Ref.      \onlinecite{Zimmer};
$^c$: Insulator-metal transition, Ref.      \onlinecite{Syassen}
\label{Ptranseu}}
\bigskip
\begin{tabular} {|l|cc|cc|}
\hline
Compound & \multicolumn{2}{c|}{$P_t$(GPa)} & \multicolumn{2}{c|}{Volume collapse (\%)} \\

      & Theory & Expt.                     & Theory & Expt. \\

\hline
EuO   & 19.3 & $30^a$, $13-30^b$           & 6.3    & $5^a$ \\
EuS   & 11.6 & $16^c$                      & 5.7    & $ 0^c$ \\
\hline
\end{tabular}

\end{table}


\widetext

\begin{table}[tbp]
\caption[b]{Calculated\cite{SSC,Act_to_be} and experimental equilbrium volumes, $V$, bulk
modulii, $B$, and $f-$electron delocalization pressure, $P$, for the
actinide elements. 
The magnetic ground state was used for Cm onwards, while the nonmagnetic was used for Pu and Am.
Experimental values are from Ref. \onlinecite{benedict},
except $^{a}$: Ref. \onlinecite{donohue}, and$^{b}$: Ref. \onlinecite{calder}%
. 
\label{act_elements}}
\bigskip
\begin{tabular}{|c|cc|cc|cc|}
\hline
& $V_{teo}$ (a.u.) & $V_{exp}$ (a.u.) & $B_{teo}$ (GPa) & $B_{exp}$ (GPa) & $%
P_{teo}$ (GPa) & $P_{exp}$ (GPa) \\ \hline
$\alpha $-Pu & 123.0(-9\%) & 135$^{a}$ &      &          &    &          \\ 
$\delta $-Pu & 163.8(-2.5\%) & 168$^{a}$ & 46.0 & 32$^{b}$ & 0 & $\sim 0$ \\ 
Am & 200.6(+1.3\%) & 198 & 46.0 & 45 & 16 & 10 \\ 
Cm & 203.8(+0.9\%) & 202 & 42.4 & 33(5) & 70 & 43 \\ 
Bk & 197.6(+6\%) & 189 & 35.0 & 25(5) & 15 & 25 \\ 
Cf & 201.8(+9\%) & 185 & 37.3 & 49(5) & 30 & 41 \\ 
Es & 256.0(-4.1\%) & 267, 321 & 19.5 & - & 11 & - \\ 
Fm & 247.4    & - & 29.6 & - &        28 & - \\
\hline
\end{tabular}
\par
\bigskip
\end{table}


\begin{thebibliography}{}
\bibitem{errorbars} The experimental transition pressures are given as
the middle of the hysteresis loop with error bars given by the extremes
of the hysteresis loop. The quoted experimental volumes at the low-
and high-pressure sides of the transition are the volumes as given by
the experimental $pV$-curves at the middle of the hysteresis loop.
\end{thebibliography}
\end{document}